\documentclass[12pt]{article}
\usepackage{epsfig}
\newcommand{\be}{\begin{equation}}
\newcommand{\ee}{\end{equation}}
\newcommand{\bea}{\begin{eqnarray}}
\newcommand{\eea}{\end{eqnarray}}
\newcommand{\ta}{\tilde\alpha}
\newcommand{\tb}{\tilde\beta}

\newcommand{\bm}{\bar\mu}
\newcommand{\bn}{\bar\nu}
\newcommand{\da}{\dot\alpha}

\newcommand{\la}{\lambda}

\newcommand{\e}{\eta}

\newcommand{\at}{\tilde a}
\newcommand{\pt}{\tilde p}
\newcommand{\omt}{\tilde \omega}
\newcommand{\om}{\omega}

\newcommand{\sss}{\sigma}
\newcommand{\ssb}{{\overline{ \sigma}}}
\newcommand{\sq}{\sqrt{2}}
\newcommand{\sqs}{\sqrt{6}}
\newcommand{\sqt}{\sqrt{3}}
\newcommand{\sqf}{\sqrt{5}}
\newcommand{\sqtt}{\sqrt{3\over 2}}
\newcommand{\os}{\overline\Sigma}
\newcommand{\s}{\Sigma}
\newcommand{\Sigb}{{\overline\Sigma}}
\newcommand{\oot}{\overline {126}}
\newcommand{\bh}{\bar h}
\newcommand{\bt}{\bar t}
\newcommand{\ovt}{\overline{10}}
\newcommand{\ovl}{\overline}
\newcommand{\nnu}{\nonumber}
\def\blfootnote{\xdef\@thefnmark{}\@footnotetext}
\begin{document}
\begin{titlepage}

\vspace{4\baselineskip}
\begin{center}{\Large\bf  MSGUTs from Germ to Bloom    : towards  Falsifiability  and Beyond
 }
\footnote{Expanded version of Plenary talks at THEP I, IIT
Roorkee,  March 16-20 and
 Planck05, ICTP, Trieste 23-28 May, 2005.}

\end{center}
\vspace{1cm}
\begin{center}
{\large
 Charanjit {S. Aulakh}$^{a, b,}$
\footnote{E-Mail:  aulakh@pu.ac.in ; Charanjit.Aulakh@cern.ch }
}
\end{center}
\vspace{0.2cm}
\begin{center}
${}^{a}$ {\it  {
Dept. of Physics, Panjab University,
Chandigarh, India 160014

\vspace{0.2cm}
${}^{b}$
 Theory Division, CERN,
CH-1213  Geneva 23,  Switzerland }}

\end{center}
\vspace{1cm}
\begin{abstract}
We review the development of    renormalizable SO(10) Susy GUTs based  on the
 ${\bf{210\oplus \oplus 126\oplus {\overline {126}}\oplus 10 }}$    Higgs system.
 These GUTs  are minimal by parameter counting.  Using the
 $SO(10) \rightarrow SU(4)\times SU(2)_L\times SU(2)_R  $
  label decomposition developed by us we calculated the complete GUT scale spectra
and couplings and the threshold effects therefrom.
  The corrections to $\alpha_G, Sin^2{\theta}_W $ and $  M_X $ are
 sensitive functions of the single parameter $\xi$ that controls
 symmetry breaking     and slow functions of
the  other parameters.   Scans of the  parameter space to identify
regions compatible with  gauge unification  are shown.

 The tight connection between the phenomenology of neutrino oscillations and
 exotic ($\Delta B\neq 0 $) processes predicted by Susy SO(10)  GUTs  is discussed in the context of
 the recent successful  fits of all available
 fermion mass/mixing data using the ${\bf{10 \oplus {\overline {126}}}}$
 Higgs representations  and  Type I/II seesaw mechanisms for neutrino mass.
 We emphasize that {\it{ the true output of these calculations
should be regarded as the unitary matrices that specify the orientation of the  embedding
of  the MSSM  within the MSGUT}}.
 The    $\Delta B\neq 0$, d=5   effective Lagrangian is written down using
these embedding matrices.
 We show how Fermion Mass fitting constraints  can be combined with GUT spectra
 to   falsify/constrain  the MSGUT and its near relatives.
 An initial survey indicates that   Type I  Seesaw  neutrino masses dominate  Type II
and both  are too small in the
perturbative MSGUT even when the mixing and mass squared  splitting ratios are  as per data.
This motivates a detailed study of the MSGUT constraints using the outputs of the
fitting of fermion data,  as well as consideration  of modifications/extensions
of the MSGUT scenario.

\end{abstract}
\end{titlepage}

\normalsize\baselineskip=15pt

\section{ Introduction}

The  Supersymmetric SO(10) GUT  based on the ${\bf 126} \oplus {\bf
{\overline{126}}} \oplus  {\bf 210}$   Higgs multiplets was first
 introduced over 20 years ago in\cite{aulmoh,ckn}. At that time
 the electroweak gauge parameters  and   fermion mass data were quite incomplete.
 The next  significant development was the realization\cite{rparityBL} that
 R parity $R_p=(-)^{3(B-L)+2S}$ formed a part of the gauge symmetry in LR
 symmetric models with gauged $U(1)_{B-L}$\cite{mohmarsh} and was hence protected by it.
 In particular it would remain unbroken as long as no field with odd $B-L$
 received a vev.  SO(10)  is now the most favoured GUT gauge group because of the natural way in which it accommodates
complete fermion families together with the superheavy right handed neutrinos  required by
the seesaw mechanism\cite{seesaw}. In 1992, with LEP data in hand and CKM parameters largely known
Babu and Mohapatra\cite{babmoh} took up the task of fitting
 Fermion masses and mixing using the ${\bf{10\oplus \oot}}$
FM Higgs system  introduced in\cite{aulmoh,ckn} and the Type I
seesaw mechanism.
 They proposed that once the charged fermion masses in the SM had been fitted
the ${\bf{10\oplus \oot}}$     matter fermion Yukawa  couplings  of the GUT would
 be completely determined  and hence the model would predict the allowed
 neutrino masses. Although, based on the then current notions about
neutrino masses/mixings,  they concluded that their proposal  failed,   it would
  inspire much future work\cite{japsnu,bsv,gohmoh,bert,babmacesnu}.

In the mid-1990's
 the  Supersymmetric LR model based on a parity odd singlet\cite{cvetic} was found to
have\cite{kuchimoh} a charge breaking vacuum and this was
construed as
 evidence of a low scale linked breaking of $SU(2)_R, B-L$ and
 R-parity\cite{kuchimoh} driven by the soft supersymmetry breaking terms.
 However an alternative analysis based on allowing this scale to lie
 anywhere in the range from $M_S$ and $M_X$  allowed the construction
of consistent Minimal Left Right Supersymmetric models
 (MSLRMs)\cite{abs,ams,amrs98} which were shown to naturally preserve
R-parity and thus predict a stable LSP. Moreover these analyses
  brought clearly to the fore a theme that had been noticed
 from the beginning\cite{aulmoh,aulphd} of the study of multiscale Susy
 GUTs : that generically in LR Susy GUTs there  are light multiplets
whose masses  are protected by supersymmetry even though they  are
 submultiplets of a  GUT multiplet that breaks gauge symmetry at a high
 scale. As such they violate  the conventional wisdom   of the ``survival  principle''
which  estimates  the masses of  such submultiplets to be the same as the large vev.
Thus it became clear\cite{abmrs99} that Susy GUTs  required the use of {\it{calculated}} not estimated
masses for RG analyses.

This ambling pace of  theoretical development was forced by the
epochal discovery of neutrino oscillations by Super-Kamiokande in 1997 and the rapid refinement of
our knowledge of the parameters thereof\cite{smirnov}.
   As is well known, the seesaw mass scale indicated by this discovery was
 in the range  indicated  by Grand Unification.
 Since LR models and GUTs containing them naturally incorporate the Seesaw mechanism,
 this gave a strong motivation for taking up the detailed study of SO(10) Susy Guts
 anew, particularly with the understanding provided by the natural class of MSLRMs
 developed earlier\cite{abs,ams,amrs98}. A completely viable R-parity preserving Susy
GUT based on the ${\bf{10\oplus \oot}}$ FM (for Fermion Mass )Higgs system and an additional
  ${\bf{45 \oplus 54\oplus 126 }}$  AM  (for Adjoint type Multiplet and euphony)
system was then developed\cite{abmrs01}.
The RG analysis carried out in this work already indicated that the use
 of calculated spectra would force together the various possible intermediate
 scales into a narrow range close to the GUT scale resulting in an effective
 ``SU(5) conspiracy''. Moreover the problem of seeing how the various MSLRMs
 considered by us would fit into Susy GUTs had motivated a review of the
 various possibilities\cite{genea,csarevs} and this had again brought up the model based
 on the 210-plet as an important and interesting possibility.

 It thus became clear that a detailed calculation of the full GUT spectrum
and couplings in various SO(10) GUT models would be necessary if
further progress was to be made and that this would require
complete calculational control on the group theory of SO(10) at
the level needed by practical field theory calculations which was
still lacking in spite of signal
contributions\cite{rabibunji,wilzee}. These techniques  --based on
an explicit decomposition of SO(10) tensor and spinor labels into
those of the maximal (``Pati-Salam'') sub-group -- were duly if
laboriously developed by us\cite{ag1}. At this time the model
based on the ${\bf 126} \oplus {\bf {\overline{126}}} \oplus  {\bf
210}$  was again brought to the fore
 by us\cite{abmsv} and its old\cite{babmoh,lee} claim to be called the minimal
Susy GUT(MSGUT) was buttressed by an analysis of its parameter
counting
 and the simplicity of its structure.

  The techniques developed by us\cite{ag1} permitted first a partial calculation
 of the mass matrices of the MSSM doublets $(1,2,\pm 1)$ and
triplets  $(3,1,\pm {2\over 3})$ \cite{ag1}  and then
 a complete calculation\cite{ag2} of all the couplings and mass matrices
of the MSGUT. With the same motivations two calculations, one in parallel
 and cross checking with ours\cite{bmsv} and another\cite{fuku04} quite
 separate, had commenced after\cite{ag1}.  These calculations both used a somewhat abstract\cite{heme}
 method that permitted the calculation of the ``clebsches'' that
 entered the spectra of MSSM submultiplets of SO(10) tensor (but not -- so far -- spinor)
 chiral supermultiplets  (but not their couplings). After some controversy concerning
 the consistency  of these three calculations
\cite{ag1,bmsv,fuku04v2,fuku0405,fuku0412,fukrebut} a consensus
seems to have emerged  on their
compatibility\cite{fuku0412,fukrebut} not withstanding notable
differences in normalizations and phase conventions.
 In\cite{ag1,ag2} we also  provided  the complete (chiral and gauge)  spectra, neutrino mass matrices,
 all gauge and chiral couplings and the effective $d=5$  operators for Baryon violation in terms
 of GUT parameters. This laid the  stage for a completely explicit RG based analysis of the MSGUT : an
 important example of which was the  calculation of threshold effects based on these
 spectra\cite{ag2} described  in Section 3.

  Meanwhile the old theme of utilizing the very restricted structure of fermion
 Higgs couplings in the class of SO(10) GUTs that used only ${\bf{10\oplus \oot}} $
representations and renormalizable couplings to make ``predictions'' concerning
the  neutrino  mass sector was taken up again. It led\cite{bsv} to a remarkably
simple (2 generation) insight into the operation of the  Type II mechanism that
 naturally generated large atmospheric  neutrino mixing angles  based on the
 observed approximate $b-\tau$ unification in the MSSM extrapolated to $M_X$.
 This simple insight then provoked a detailed analysis of the fitting problem
for dominant Type II seesaw mechanism and  3
generations\cite{gohmoh}
 which was  quite  successful.
 A certain tension in the details of these fits was
ameliorated by later work\cite{bert}  and disappeared\cite{bert}
when a {\bf{120}}-plet was also allowed to perturb the fit of the
${\bf{10 + \oot}}$ slightly.
 Finally, very recently, in fact just in time for  PLANCK05 Babu announced
 that   they \cite{babmacesnu} had succeeded in obtaining viable Type I,
Type II and mixed fits to the (large mixing angle) neutrino mass data: which possibility had   been
 neglected since the suggestion of\cite{bsv}. These sudden reversals are
  due to the extreme delicacy and complexity of the  task of
 fitting accurately the masses of particles differing by a factor of upto
$10^6$ in a multidimensional complex parameter space where
algorithms ``shoot''  obliviously past solution points.

 Obviously these successful fits of all low energy fermion ``quadratic'' data
(extrapolated up to $M_X$ ) in generic MSGUTs cry out to be
married to a particular simple MSGUT in which all coefficients are
specified and testable
 for viability. This is particularly so because only in a particular MSGUT,
 where the needed coefficients are fully specified\cite{ag2} can one test
the viability of  a given FM fit (such as the Type II dominant
ones\cite{bsv,gohmoh,bert}). Furthermore, as we shall see in
detail below,{\it{ the true output of the FM fit is
 really the embedding matrices that specify the relation between MSSM and MSGUT
 supermultiplets  at the GUT scale.}} Given the phenomenologically expected quasi-identity of flavour
 mixing in the scalar and fermionic sector down to    Electroweak  scales, these
 matrices are the  main  missing ingredient for performing an informed and  realistic calculation of Baryon decay
 and Lepton Flavour violation in the MSGUT. {\it{The combined  cuts provided by gauge unification and
fermion mass fit viability    constrain
  the MSGUT  and render it falsifiable }}.  If  it passed these tests with some region of its
 parameter space unscathed we would be in a position to actually specify the
 theory {\it{at the high scale}}: a task long thought to be impossible by
 those despondent at the continuing failure to detect proton decay. The key
 to this remarkable development is of course the remarkable uroborotic
(  $o  \upsilon\rho o   \beta o \rho o \varsigma $
:  the  world-snake that swallows its tail) feature of
the Seesaw which connects the physics of the ultralight and  ultramassive
 particles.

   In Section II we review the structure and spontaneous symmetry breaking
 of the MSGUT and describe its decomposition to Pati-Salam labels. This
is  accompanied by  an Appendix  containing  the complete spectrum of MSSM
 multiplets in the MSGUT -including gauge submultiplets. In Section III we
 describe the  RG analysis of threshold effects and the type of cuts these
 corrections impose on the parameter space.
   In Section IV we review the fitting of Fermion masses and mixings with a
 view to supporting our assertion that the true output of this procedure
 is the specification of the embedding matrices. In Section V we plot the
 behaviour of the parameters that control Type I vs Type II dominance in
the MSGUT and show that Type II dominance may be quite unlikely. We also use
 particular examples of solutions provided  by one of the groups\cite{bert} that have
performed the successful fits\cite{gohmoh,bert,babmacesnu} to plot
the Type I and Type II contributions and again\cite{gohmoh04}
find  indications that Type II is not
  easily  dominant and that the overall size of neutrino masses tends
to be too low except possibly in very special regions of parameter space.
In the next section we transform our previously
 derived formula for the effective $d=5$  Lagrangian for Baryon decay in the
MSGUT to the MSSM  using the embedding matrices. This operator is then
 ready for use to calculate Baryon decay.  We also briefly comment on the relevance of our considerations
to $d=6$ and Lepton Flavour violation. We conclude with a
discussion of the outlook and the directions that will be pursued
in the forthcoming detailed paper\cite{newtas} on the
 application of the techniques indicated here to the pinning down of the MSGUT.

\section{Essentials of  MSGUTs}

There are at present two main types of SO(10) Susy GUT models
 namely renormalizable GUTs (RGUTs)\cite{aulmoh,ckn,abmrs01} of
the classic type which invoke only gauge symmetry  and  preserve R-parity by
 maintaining the sharp distinction between Matter and Higgs Supermultiplets,
 and  non renormalizable GUTs (nRGUTs)\cite{nRGUTs}
  which allow non renormalizable operators - particularly for
generating fermion masses - but  impose additional symmetries to
perform the functions of R-parity and suppress unwanted behaviours
allowed by the license granted by non renormalizability.
 Practitioners of the art are often  sharply fixed
 in their preference for one type or the other and our own is obviously
 with the former.  Many (talks by Babu,Malinsky,Mohapatra..)  of
the contributions on GUTs to   PLANCK05  are of this type but the
 other type is also well represented   (Pati,Raby...).

Other distinguishing features of these two types of models  are that
 RGUTs  allow large representations such as the ${\bf{{\overline {126 }}}}$ Higgs
to generate charged and neutral (via type I and Type II Seesaw)
 fermion masses but claim minimality on grounds of   maximal
 economy of parameters. In contrast  nRGUTs, allow only
 small Higgs  representations and use non renormalizable
operators(${\bf{16_F^2 {\overline{16}}_H^2 }}$ etc)  to generate effective
vevs in the ${\bf{ \oot,   120}}$ channels. They achieve simplifications
by invoking additional symmetries, whose rationale is  however not always
appreciated  by non adherents.  A very large number of works on
 nRGUTs\cite{nRGUTs} have appeared and some of these have   far   advanced
  fitting programs :  with claims to generate sample spectra with
specific numbers for most (i.e  $\sim 10^2$) low energy MSSM parameters. Moreover they also
seek to satisfy additional motivations like workable models of gauge
 mediated supersymmetry breaking\cite{nRGUTs}.

The RGUTs, on the other hand are mostly at the stage of seeking a
semi quantitative non-disagreement with the data as available at
present. For example, since no superpartner has as yet  actually
been observed,
 and speculations on their possible masses range all the way from the
 classic 1 TeV up to almost the GUT scale, the value of the low energy
 parameters renormalized up to the Susy breaking scale where the Susy
 RG  equations take hold, can, realistically, be considered as known
 to at  best an accuracy of $5-10\%$. Invoking the remarkable accuracy
 of experimental data at the scale $M_Z$ is of little avail since
 nothing is known of the size of the threshold corrections due to superpartners.
It is this approach that we shall adopt in this paper and our
quantitative analyses regarding quantities unprotected by any symmetry
shall not lay claim to accuracy that may turn out to be quite spurious.

         A very important but controversial distinction   between the
 two  classes of theories is due to  the fact that the  one loop beta
functions of the $ {\bf{\oot, 120}}$  chiral supermultiplets  are so large (35 and 28
 respectively)  that they alone overcome the contribution (-24) of the
 SO(10) gauge supermultiplet and cause the gauge coupling to explode just
above $M_X$ the perturbative unification scale.  By analyzing the
complete two loop RG equations for the MSGUT above $M_X$ we have
shown\cite{RGSO(10)} that the large positive coefficient in the
one loop gauge  beta function precludes evasion of this difficulty
by taking refuge in a weakly coupled fixed point before the gauge
coupling explodes.
 {\it{Thus such theories effectively determine their own upper cutoff
 $\Lambda_X \sim 10^{17} GeV $ }}. This feature is so inescapable
 that  we proposed\cite{trmin,tas} to face it by making it a signal of a
deeper non-perturbative feature of such Susy GUTs: namely that they
 dynamically break the GUT symmetry down to a smaller symmetry (such
 as $H=SU(5) \times U(1)$  or even  $H=G_{123}$ ) at the scale $\Lambda_X $ via a
 {\it{Supersymmetric }} condensation of gauginos in the coset
 $SO(10)/H$  which drives a $G_{123}$ preserving Chiral scalar
 condensate\cite{tas}.  A new fundamental length $\Lambda_X^{-1} $  below which SM probes
 cannot  delve thus arises, endowing the particles of the SM with ``hearts''
 i.e impenetrable cores with sizes $\sim 10^{-30} cm $.   We have
 implemented\cite{tas} the strongly coupled Supersymmetric dynamics
 required by this  type of scenario in a toy model based on $SU(2)$
 which can be easily generalized to at least the simple breaking
$SO(10) \longrightarrow SU(5) \times U(1) $\cite{newtas}.
 Various fascinating perspectives including dual unification
and induced gravity then beckon. However in view of the
controversial nature of this proposal we shall not  treat of
 it further here.  This talk is aimed at indicating the
feasibility of a promotion of  the MSGUT to a  falsifiable
 theory using the deep linkages between low and high scale
physics inherent in the SO(10) seesaw mechanism  in combination with the traditional -relatively
 permissive- constraints of gauge unification.

 The MSGUT is the renormalizable  globally supersymmetric $SO(10)$ GUT
 whose chiral supermultiplets  consist of ``AM type''  totally antisymmetric tensors : $
{\bf{210}}(\Phi_{ijkl})$  $   {\bf{\overline{126}}}({\bf{\Sigb}}_{ijklm}),$
 ${\bf{126}} ({\bf\Sigma}_{ijklm})(i,j=1...10)$ which   break the GUT symmetry
 to the MSSM, together with Fermion mass (FM) Higgs {\bf{10}}-plet(${\bf{H}}_i$).
  The  ${\bf{\overline{126}}}$ plays a dual or AM-FM
role since  it also enables the generation of realistic charged
fermion   and    neutrino masses and mixings (via the Type I and/or
Type II mechanisms);  three  {\bf{16}}-plets ${\bf{\Psi}_A}(A=1,2,3)$  contain the
matter  including the three conjugate neutrinos (${\bar\nu_L^A}$).

If in addition to the ${\bf{10,\oot}}$ FM Higgs  we also include a ${\bf{120}}$ -plet
 Higgs allowed  by the SO(10) multiplication rule ${\bf{16\times 16 = 10 + 120 + 126 }}$, the
 GUT scale symmetry breaking is unchanged since the ${\bf{120 }}$ contains no SM
 singlets.   However it does contribute {\it{two}} additional pairs of $(1,2,\pm
 1)$ doublets taking the MSSM Higgs doublet mass matrix ${\cal H}$ from
 $4\times 4$ to $6\times 6 $. The resulting theory  is thus justly
 called the next to minimal Susy GUT (nMSGUT).

 The   superpotential   is
 \bea
  W &=&{1\over{2}}M_{H}H^{2}_{i} + {m \over
{ 4! }} \Phi_{ijkl}\Phi_{ijkl}+{\lambda \over
{{4!}}}\Phi_{ijkl}\Phi_{klmn} \Phi_{mnij}+{M \over { 5!
}}\Sigma_{ijklm}\overline\Sigma_{ijklm}\nonumber\\
 &+&{\eta \over
4!}\Phi_{ijkl}\Sigma_{ijmno}\overline\Sigma_{klmno}+
 {1\over{4!}}H_{i}\Phi_{jklm}(\gamma\Sigma_{ijklm}+
\overline{\gamma}\overline{\Sigma}_{ijklm}) \nonumber \\
  &+& h_{AB}'\psi^{T}_{A}C^{(5)}_{2}\gamma_{i}\psi_{B}H_{i}+ {1\over
5!}f_{AB}' \psi^{T}_{A}C^{(5)}_{2}{\gamma_{i_{1}}}...
{\gamma_{i_5}}\psi_{B}\overline{\Sigma}_{i_1...i_5}
 \label{W}
 \eea
  The parameter  counting is as follows : the  7  complex parameters
$m,M,M_H,\lambda,\eta,\gamma$ and ${\bar{\gamma}}$  can be relieved of
 4 phases using the arbitrariness in the phases of the 4 SO(10)
 multiplets  ${\bf{210,126,\oot,10}}$ leaving 10 real parameters. The 12 complex
 Yukawa contained in the symmetric $3\times 3$  matrices $h',f'$
can be reduced to 15 real parameters by diagonalizing one linear
 combination of the two to a real diagonal form.   In addition there
 is the gauge coupling.
In all the MSGUT thus has exactly 26 non-soft parameters
\cite{abmsv}.
 Incidentally the  MSSM also has 26 non-soft couplings.
 In\cite{abmsv} we have shown that this number of ``hard'' parameters
is considerably less than any GUT attempting to perform the same
tasks vis a vis fermion masses that the MSGUT accomplishes. The
`levity of the  cognoscenti' provoked by the number 26 will recur
again below ! Note however that one of  these parameters, e.g
$M_H$,   is traded for the electroweak vev via the fine tuning
condition that yields two light doublets and another  Susy
electro-weak
  parameter i.e $\tan \beta$ must be taken
as an additional  given of the analysis since it is likely
determined by dynamics dependent on the supersymmetry breaking
parameters.

The GUT scale vevs that break the gauge symmetry down to the SM
symmetry (in the notation $a,b=1...6 ; {\tilde\alpha}=7...10 $ of
\cite{ag1})  are
 {\cite{aulmoh,ckn}}
${\langle(15,1,1)\rangle}_{210} : \langle{\phi_{abcd}}\rangle={a\over{2}}
\epsilon_{abcdef}\epsilon_{ef} ;$\hfil\break $
\langle(15,1,3)\rangle_{210}~:~\langle\phi_{ab\ta\tb}\rangle=\omega
\epsilon_{ab}\epsilon_{\ta\tb}\quad ;\quad
  \langle(1,1,1)\rangle_{210}~: ~\langle\phi_{ {\tilde
\alpha}{\tilde \beta} {\tilde \gamma}{\tilde \delta}}
\rangle=p\epsilon_{{\tilde \alpha} {\tilde \beta} {\tilde
\gamma}{\tilde \delta}}\quad; $\hfil\break $
\langle(10,1,3)\rangle_{\oot} ~:
  \langle{\overline\Sigma}_{\hat{1}\hat{3}\hat{5}
\hat{8}\hat{0}}\rangle= \bar\sigma \quad;
\langle({\overline{10}},1,3)\rangle_{126} ~:
\langle{\Sigma}_{\hat{2}\hat{4}\hat{6}\hat{7}\hat{9}} \rangle=\sigma $.
 The vanishing of the D-terms of the SO(10) gauge sector
 potential imposes only the condition $
 |\sigma|=|{\overline{\sigma}}| $.
Except for the simpler cases corresponding to enhanced unbroken
symmetry  ($SU(5)\times U(1), SU(5)$\hfil\break $  G_{3,2,2,B-L},
G_{3,2,R,B-L}$ etc)\cite{abmsv,bmsv} this system  of equations is
essentially cubic and can be reduced to  the single
equation\cite{abmsv}
 for a variable $x= -\lambda\omega/m$, in terms of
 which the vevs $a,\omega,p,\sigma,
 {\overline\sigma}$ are specified  :

\be 8 x^3 - 15 x^2 + 14 x -3 = -\xi (1-x)^2 \label{cubic} \ee

where  $\xi ={{ \lambda M}\over {\eta m}} $.  This   exhibits the
crucial importance of the parameter $\xi$.

Using the above vevs and the methods of\cite{ag1} we calculated
the complete   gauge and chiral multiplet GUT scale spectra
{\it{and}} couplings for the 52 different MSSM multiplet sets
falling into 26 different MSSM multiplet types of which 18 are
unmixed while the other 8 types occur in multiple copies. ({\it{On
the  lighter note}} : the occurrences yet again of the `mystic'
String Theory number 26
 demonstrates that one can do just as well without string theory  when searching for the fundamental
theory !). These spectra may be found in   the Appendix.

Among the mass matrices exhibited is the all important $4\times 4$
Higgs doublet mass matrix\cite{ag1,ag2} ${\cal H}$. To keep one
pair of these doublets light one  tunes $M_H$ so that $Det{\cal
H}=0$. This matrix can then be  diagonalized  by a bi-unitary
transformation yielding thereby the coefficients describing the
proportion of the doublet fields in ${\bf{10,\oot,126,210}}$
present in the light doublets : which proportions are important
for many phenomena.

 \section{ RG Analysis}

If the serendipity\cite{marsenj,jones,amaldi} of the MSSM gauge coupling unification at $M_X^0$
 is to survive closer examination the  MSGUT  must  answer the query :

 {\it{ Are the one loop values of $Sin^2\theta_W $ and
 $M_X$ generically stable against superheavy threshold
corrections ?}}.

   We follow the approach of Hall\cite{hall} in which the mass of the lightest baryon number violating
superheavy gauge bosons   is chosen    as the common   ``physical" superheavy
matching point  ($M_i=M_X$)  in the equations relating
  the   MSSM couplings to the SO(10)   coupling\cite{hall}  :
\bea {1\over{\alpha_i(M_S)}}
 ={1\over{\alpha_G(M_X)}} +
 8 \pi b_i ln{{M_X}\over{M_S} }
  + 4 \pi \sum_j {{{b_{ij}} \over {b_j}}} ln X_j -
4\pi\lambda_i(M_X) \eea See\cite{ag2} for details.   In this
approach, rather than enforcing  unification at a point, it is
recognized  that
 above the  scale $M_X$ the effective theory changes to a Susy  $SO(10)$ model {\it{structured by the
complex superheavy spectra which we have computed }}  and which
appears as unbroken SO(10) only at the crudest resolution -- here
surpassed.   Thus we compute the corrections to  the three
parameters $Log_{10}{M_X}, sin^2\theta_W (M_S), \alpha_G^{-1}(M_X)
$
  as a function of the MSGUT  parameters  and the answer to the question of stability of the
 perturbative unification is determined by  the ranges of GUT parameters where these
corrections are consistent with
the known or surmised data on  $Log_{10}{M_X}, sin^2\theta_W (M_S) $ and the consistency requirement that
the   SO(10) theory remain perturbative after correction. We find the corrections
\bea
 \Delta^{(th)}(Log_{10}{M_X})  &=&.0217 +.0167 (5 {{\bar b}'}_1 +3{{\bar b}'}_2 -8
  {{\bar b}'}_3) Log_{10}{{M'}\over  {M_X^0}} \label{Deltasw}\\
\Delta^{(th)} (sin^2\theta_W (M_S)) &=&
 .00004 -.00024 (4 {{\bar b}'}_1 -9.6 {{\bar b}'}_2 +5.6
  {{\bar b}'}_3) Log_{10}{{M'}\over  {M_X^0}}
  \label{Deltath}\\
\Delta^{(th)} (\alpha_G^{-1}(M_X)) &=&.1565 +.01832 (5 {{\bar
b}'}_1 + 3 {{\bar b}'}_2 + 12 {{\bar b}'}_3) Log_{10}{{M'}\over
{M_X^0}}
  \eea
Where ${\bar b'}_i = 16\pi^2 b_i'  $ are   1-loop
$\beta$ function coefficients ($ \beta_i=b_i g_i^3 $)  for multiplets with
 mass $M'$ (a sum over representations is implicit).

These corrections are to be added to the one loop values
corresponding to the successful gauge unification of the MSSM  :

Using the values
  \bea
    \alpha_G^0(M_X)^{-1}= 25.6\quad ; \quad M_X^0=10^{16.25} ~GeV \quad ;\quad
 M_S=1 ~  TeV \nonumber \\
 \alpha_1^{-1}(M_S)=57.45 \quad ;\quad  \alpha_2^{-1}(M_S)=30.8 \quad ;
  \quad\alpha_3^{-1}(M_S)=11.04
 \eea
 the two loop corrections are

\bea
 \Delta^{2-loop}(log_{10}{{M_X}\over {M_S}})&=&-.08 \qquad ;
  \quad \Delta^{2-loop}(sin^2\theta_W (M_S))=.0026
\nnu\\
 \Delta^{2-loop}\alpha_G^{-1}(M_X)&=& -.546
  \eea

We see that in comparison with the large threshold effects to be
 expected  in view of the number of heavy fields and their beta
 functions\cite{dixitsher,ag2}  the 2 loop corrections are quite small.

A few remarks on the role of the parameters  are in order.
The parameter $\xi= \lambda M/ \eta m$ is the only
  numerical parameter that  enters into the cubic  eqn.(\ref{cubic})
 that determines the parameter $x$ in terms of which all the  superheavy vevs are given.
{\it{ It is thus
 the most crucial  determinant of the mass spectrum }}.  The dependence
of the threshold corrections on the parameters
 ${\lambda,\eta,\gamma,{\bar\gamma}}$  is comparatively mild  except
when coherent e.g when  many masses are lowered  together
leading to $\alpha_G $ explosion.
     The parameter  ratio   $m/\lambda$ can be extracted as the overall scale
  of the vevs.   Since the threshold corrections    we calculate are dependent  only on (logarithms of)
  ratios of masses,   the parameter $m$ does not play any crucial role in our
  scan of the parameter space : it is simply fixed   in terms of the
  mass $M_V=M_X$   of the lightest superheavy vector particles mediating
  proton decay.

With these formulae in hand we can explore the dependence of the
 threshold corrections on the ``fast'' parameter $\xi$ in response to
 which the vevs and thus all the threshold corrections can gyrate
 wildly (see plots  for the real solution for real $\xi$ below ),
 and the ``slow diagonal ''  parameters $\lambda,\eta$  whose
lowering tends to make fields light and thus give large negative
 corrections to $\alpha_G^{-1}$. In addition there are the
 ``slow off diagonal'' parameters $\gamma, \bar\gamma$ whose
 effect  seems  quite  mild.

$ Sin^2\theta_W $ is now very accurately known\cite{pdg} at $M_Z$:
${\hat s}_Z^2=.23120 \pm.00015$ or an error of less than $.065
\%$. Similarly the value  $\alpha_{em} = 127.906\pm.0019$ has an
error of only $.015\%$. The main uncertainty in the data (besides
$M_H$  )   at $M_Z$ is in $\alpha_3(M_Z)$ : $ \sim 1.5\%$. However
the same is obviously not true of the overall sparticle mass scale
or the intra  sparticle mass splittings : with values in current
speculation ranging from $\sim 1$ TeV    to $ 10^{10} $ TeV  !
Thus at least until the first superpartners are observed the
assumption that we know the values of the $\alpha_i(M_S)$ with
anything like the precision at $M_Z$ is quite unjustified. A rough
estimate of the uncertainty in $Sin^2\theta_W(M_S) $ gives numbers
in the ball park of 1\% to 10\% and we shall not pretend to have
any license to impose stronger constraints on our parameters.  For
the uncertainty in $Log_{10} M_X$ however there is a quite
stringent $M_S$ independent  constraint coming from the
requirement that gauge mediated proton decay in channels like
$p\rightarrow e^+ \pi^0$ obey the current bounds
$\tau_{p\rightarrow e^+ \pi^0 } > 10^{33} yr$. Thus $\Delta
Log_{10} M_X > -1 $ is a very firm constraint that we may rely
upon. Finally it is clear that the applicability of the
perturbative formulae used in our treatment will need detailed
scrutiny  if the fractional change in $\alpha_G(M_X)^{-1}$ is
greater than about 25\%(particularly if the change is negative).
 So  $|\Delta \alpha_G(M_X)^{-1}|<10 $(say) is a limitation we  tentatively observe as a limit on the
range of validity  of our calculation (  rather than the theory
itself : since a detailed examination may permit one to handle
larger changes  -- particularly positive ones --  in
$\alpha_G(M_X)^{-1}$). In view of these relatively loose bounds on
the acceptable changes associated with threshold effects we cannot
expect that gauge constraints can limit the allowed parameter
space very precisely. Nevertheless, these constraints used in
conjunction   with additional very significant information arising
from the fitting of Fermion masses and (in more model dependent
ways) $d=5$ Baryon violation and Lepton flavour violation could
allow us to obtain a complete   semi-quantitative  picture of
viable regions, if any, of  the MSGUT parameter space.

    We now present examples showing how such a semi-quantitative mapping
 of the RG constrained topography of the GUT scale parameter space
may be obtained using the formulae given above. An attempt at an
exhaustive mapping  would be premature till the    recent available  FM fits
\cite{bert,babmacesnu} have been digested.  Later we shall present our
 program of further constraining the parameter space using the
constraints of the  fit\cite{bsv,gohmoh,bert,babmacesnu}
 of  Fermion masses and also due to the consequent specification
 of Baryon decay operators.     It should be kept
in mind that at any $\xi$ the cubic equation (\ref{cubic} ) has three
 solutions  any of which is in principle exploitable for defining a
 vacuum.

   Consider first  the plots of the threshold  changes
 $ \Delta^{(th)}  (\alpha_G^{-1 }(M_X))$  and  $ \Delta^{(th)} (Log_{10}{M_X})$ vs $\xi$
after (arbitrarily) setting
  $\{\lambda,\eta,\gamma,\bar\gamma\}=\{.7,.5,.3,.2\}$ which are shown as Fig. 1 for the real solution
of the eqn(\ref{cubic}) for real $\xi$.

 The    ``twin towers'' due to singularities at
 $\xi=-5,10$  in the plot of   $\alpha_G(M_X)^{-1}$ arise from
the $SU(5) $ invariant real vevs at these
 values of $\xi$. Similarly the negative spike at $\xi=-2/3$
 corresponds to a real  solution with $SU(5)_{flipped} \times U(1) $
symmetry\cite{abmsv,bmsv}.  As emphasized by us in\cite{ag2} the
 plots of  the threshold corrections show very sharply defined  features
 corresponding  to the special behaviour near points of enhanced symmetry and thus
 provide  a rapid way of scanning the topography of the parameter space. For
 these  moderately large values of $\lambda,\eta$  the lower cut at
 $\Delta\alpha_G(M_X)^{-1}=-10$   essentially allows all possible
 values of $\xi$ though the upper cut  does advise
 caution around the special symmetry points with $SU(5)$
 symmetry. However  the graph of $\Delta Log_{10} M_X$ with a cut at
$\Delta Log_{10} M_X=-1 $ immediately rules out most of the region
 $\xi\in(.25, 8.6)$  due to the ultrarapid gauge boson mediated
Proton decay in that range of $\xi$.

When we lower the value of $\lambda$ or $\eta$  many particles
became light so that  the entire $\alpha_G(M_X)^{-1}$  vs $ \xi$
plot shifts downward. This is seen in Fig. 2 where we repeat the
same Plot as Fig 1. but now with $\lambda=.1 $. As a result
{\it{larger}} $\xi$ values are allowed  on grounds of limiting the
change in $\alpha_G(M_X)^{-1}$.   On the other hand  the behaviour
of   $\Delta Log_{10} M_X$ in response to a decrease in the
diagonal slow parameters $\lambda,\eta$  is quite mild.   Thus
although lowering $\lambda$ does tend to make $\Delta Log_{10} M_X
$ less negative in $\xi>0$ region, the large positive change in
$\alpha_G(M_X)$  would rule out the small $\lambda,\eta$  region.
The condition $\Delta Log_{10} M_X  >-1$ required by proton
stability rules out the region $.25 <\xi < 8.6$ for the real case
(see Figs. 1., 2.).

 Next  we plot  $\Delta Sin^2\theta_W(M_S)$ versus $\xi$ for
moderate and very  small $\lambda$   (Fig 3.).  The change in
$\lambda$ has practically no effect. We also  see that there are
ranges of $\xi$ where the change in $  Sin^2\theta_W(M_S)$  is
less that $ 10 \%$ and that these ranges are  only slightly
affected by the change in $\lambda$. However the region
$2.8<\xi<8$  which was excluded by  $\Delta Log_{10} M_X  >-1$  is
also excluded by the large change
 $\Delta Sin^2\theta_W(M_S)$.

The further evolution of $\Delta^{(th)} (\alpha_G^{-1}(M_X))$  and
$\Delta^{(th)}(Log_{10}{M_X})$ as $\lambda$ is decreased to $.01$
is similar as can be seen in Fig.4.  Lowering $\eta$ has effects
very similar to those of lowering $\lambda$ since the effect of
smaller values for these diagonal couplings is to lower the masses
of whole sets of multiplets and therefore raise $\alpha_G (M_X)$
sharply.  The effects of the off diagonal couplings $\gamma$ and
$\bar\gamma$ are  much milder.

 With $\xi$ real there are also two complex (mutually conjugate) solutions of the basic cubic
equation (\ref{cubic}). Examples of these are displayed as Fig. 5,6  for moderate  and small
 $\lambda$.
 We observe that the  behaviour of the complex solution  is   smoother than
 the real one and that
 apart from  the spikes observed at $\xi=-5$ there is hardly any sharp feature to be seen.
 The effect of decreasing $\lambda$ or $\eta$  on $\Delta^{(th)} (\alpha_G^{-1}(M_X))$
is however even stronger than in the case of real  $x$ as may be
seen in Fig 6,  thus restricting the viable magnitudes of
$\lambda,\eta$ to moderately large values again. The corrections
to $Sin^2\theta_W (M_S)$ for the complex solution are shown
 for moderate and very small values of $\lambda $ as Fig. 7.
 One can also consider complex values of the parameter $\xi$ and the
three solutions in that case. The behaviour is quite similar to
that shown for the complex solutions for real $\xi $  but needs
detailed and comprehensive examination\cite{newtas}.

 We have exhibited these graphs to give a sense of the
structure  visible once one ramps up the resolution of  analysis
to reveal the  finestructure hidden within the bland
impressiveness of ``supersymmetric unification at a point''. As
already noted long since\cite{barbhall} the ambiguities associated
with superheavy thresholds  would  not allow one to predict the
effective scale of superpartner masses or the unification scale
even if the low energy values $\alpha_i(M_Z)$  were known exactly.
In fact,
 following Hall\cite{hall}
we have chosen not to treat the unification beyond leading order by imposing unification
at a point
but rather in terms of quantifying the ambiguities in $  \alpha_G^{-1}(M_X) $,
$Sin^2\theta_W (M_S) $    and  $  Log_{10}{M_X} $ caused by the  finestructure
of unification scale mass spectra.  The range of behaviours exhibited make it
unlikely that the constraints of gauge unification alone will rule out this
minimal Susy SO(10) GUT or fix its parameters. However when taken
together with the rest of the low energy data,  the MSGUT provides a
 well defined and calculable framework within which
significant questions regarding the viable GUT scale structures
can be posed and answered. Furthermore  the
process\cite{bsv,gohmoh,bert,babmacesnu} of fitting the highly
structured  fermion data to the relatively few parameters of the
MSGUT  excavates crucially important information regarding the
embedding of the MSSM within the MSGUT. This information
(concerning mixing angles of various sorts) is critical for
determining the precise predictions of the MSGUT for both $d=5, 6$
baryon number violating operators in the effective MSSM as well as
the predictions for Lepton Flavour violation.

\section{FM Fitting Frenzy}

 We have already reviewed the sequence of developments preceding the current focus of
interest on the fitting of fermion mass and mixing data in the
MSGUT\cite{babmoh,japsnu,bsv, bert,babmacesnu}. The fitting
program itself has  used only the {\it{form}} of the fermion mass
formulae in the MSGUT (which follows from the use of only the
${\bf{10 + \oot}}$ representations) rather than the specific
formulae for  the coefficients in the fermion masses  dictated by
the MSGUT superpotential. Our concern here is  not with  the
actual values of  the successful
 fits but rather their implications when combined with the structure of the MSGUT.
We therefore review the fitting procedure  from our particular
viewpoint. To begin with the ``Clebsch ''  coefficients for the
couplings of the ${\bf{16 \times 16}}$ SO(10) chiral spinors to
the ${\bf{10,\oot}}$ irreps were calculated as a part of our
explicit decomposition of $SO(10)$ in terms of Pati-Salam
labels\cite{ag1} :  One obtains : \bea W_{FM}^H &=& h_{AB}'
\psi^{T}_{A}{C}_{2}^{(5)}
\gamma_{i}^{(5)}\psi_{B} H_{i} \nonumber\\
 &=& \sqrt{2}h_{AB}' \big [H_{\mu\nu}\widehat{\psi}^{\mu\da}_{A}
\widehat{\psi}^{\nu}_{B\da} +\widetilde{H}^{\mu\nu}\psi_{\mu
A}^{\alpha} \psi_{\nu\alpha B}
 - H^{\alpha\dot\alpha}
(\widehat{\psi}^{\mu} _{A\dot\alpha}\psi_{\alpha\mu B}
+\psi_{\alpha\mu A}\widehat{\psi}
_{\dot\alpha B}^{\mu} )\big ]\nonumber\\
&=&-2\sqrt{2}h_{AB}'\bar{h}_{1}[{\bar d}_{A} Q_{B} +\bar
e_{A}L_{B}] +2\sqrt{2}h_{AB}'{h}_{1}\big [{\bar u}_{A} Q_{B}
+{\bar\nu}_{A}L_B \big]\nnu\\
&+&..... \eea
 \bea   W_{FM}^{\Sigb} &=&   {1 \over 5!}
\psi^{T}C_{2}^{(5)}\gamma_{i_{1}}.....\gamma_{i_{5}}\chi
{\overline\Sigma}_{i_{1}...{i_{5}}} \nonumber\\
\nnu\\
&=& {4 \sqrt {2}}{\overline\Sigma}^{~\mu\alpha\dot\alpha}_{\nu}
 {(\widehat\psi_{\dot\alpha}^{\nu}\chi_{\alpha\mu}+\psi_{\mu\alpha}\widehat
\chi_{\dot\alpha}^{\nu})}
+ 4({\overline\Sigma}_{\mu\nu}^{\dot\alpha\dot\beta}\widehat\psi_{\dot\alpha}^{\mu}
\widehat\chi_{\dot\beta}^{\nu}+{{\overline\Sigma}}^{\mu\nu\alpha\beta}\psi_{\mu\alpha}
\chi_{\nu\beta})..... \nonumber\\
    &=&   4{\sqrt{2}}f_{AB}'
 [{i\over\sqrt{3}} \{ \bar{h}_{2}({\bar
d}_{A}Q_{B}- 3{\bar e}_{A} L_{B})
 -{h}_{2}({\bar u}_{A}Q_{B}
-3{\bar\nu}_{A}L_{B})\}
  \nonumber\\
 &+& 4 f_{AB}' [ -2 i G_{5}{\bar\nu}_A{\bar\nu}_B)
+\sq {\ovl O} L_{A}L_{B})\nonumber +......\\
 \label{MMFMS} \eea

where the alphabetical  naming convention regarding the
subcomponents of the Higgs multiplets is given in the Appendix and
in  detail in\cite{ag2}. From the properties of the SO(10)
Clifford algebra it follows that the Yukawa coupling matrices
 ${ h'}_{AB},{ f'}_{AB}( A,B=1,2,3 )$ are {\it{symmetric}}
(${ h'}_{AB}= { h'}_{BA},{ f'}_{AB} =  { f'}_{BA}$) complex
matrices.
 Therefore the freedom to  make unitary changes of  basis
 allows one to chose one linear combination
of the matrices ${ h'},{ f'}$ to be diagonal. We shall choose the
basis where $ f'$ is diagonal since it proves convenient when
analyzing the Seesaw mechanism
 but our conclusions are independent of any such choice.

To obtain the formulae for the charged fermion masses from the
above decomposition\cite{ag2} one first needs to define the
($G_{321} (1,2,\pm 1) $)  multiplets $H^{(1)},{\bar H}^{(1)}$
which are the (light) MSSM Higgs doublet pair. This is  achieved
by imposing the condition

  $$\quad Det {\cal H}  \sim O(M_W)$$

on the doublet mass matrix ${\cal H}$ which occurs in the quadratic terms
of the superpotential when expanding around the superheavy vevs:
 $W={\bar h} {\cal H} h +... $. This amounts to a fine tuning of(say)  the mass parameter $M_H$
of the 10-plet Higgs.

The $4\times 4$ matrix ${\cal H}$\cite{ag1}  can be diagonalized
by a bi-unitary transformation : \cite{abmsv,bmsv,ag1,ag2} from
the 4 pairs of Higgs doublets $h^{(i)},{\bar h}^{(i)}$ arising
from the SO(10) fields to a new set $H^{(i)},{\bar H}^{(i)}$ of
fields in terms of which the doublet mass terms  are diagonal.

\bea {\overline U}^T {\cal H}U &=&   Diag ( m_H^{(1)},m_H^{(2)},....)
 \nnu\\
 h^{(i)} &=& U_{ij} H^{(j)}  \qquad ;\qquad   {\bar h}^{(i)} = {\bar
U}_{ij} {\bar H}^{(j)}  \eea

 Then it is clear that in the effective theory at low energies the  GUT Higgs doublets
 $h^{(i)},{\bar h}^{(i)}$  are present in   $H^{(1)},{\bar H}^{(1)}$ in a
 proportion determined by the
 first columns of the matrices  $U,{\bar U}$ :

    \bea E < <M_X \qquad : \qquad  {  h}^{(i)} &\rightarrow&  {  \alpha}_i {  H}^{(1)} \quad
  ; \quad
{ \alpha}_i = {  U}_{i1} \nnu\\
   {\bar h}^{(i)} &\rightarrow&  {\bar \alpha}_i {\bar H}^{(1)} \quad
  ; \quad
{\bar\alpha}_i = {\bar U}_{i1}
  \eea

Thus the formulae for the charged fermion masses in the MSGUT
are\cite{ag2} : \bea
 M^d &=& {  r_1} {\hat h} + { r_2} {\hat f}   \nnu  \\
  M^l &=&{  r_1} {\hat h} - 3 {  r_2} {\hat f}  \\
 M^u &=&   {\hat h} + {\hat f}   \nnu\eea
where
 \bea
 \hat{h} &=& 2\sqrt{2}h^{\prime} v \alpha_1 \sin\beta\nnu\\
 \hat{f} &=& -4\sqrt{\frac{2}{3}} i f^{\prime}{\alpha_2}
 \sin\beta  \\
 r_1 &=& \frac{\bar{\alpha_1}}{\alpha_1}\cot\beta \nnu\\
  r_2  &=&
 \frac{\bar{\alpha_2}}{\alpha_2}\cot\beta
 \nnu \eea

here ($h^{\prime},f^{\prime})$ are the couplings in the MSGUT superpotential.

Similarly the Majorana mass of the `right handed neutrinos ' i.e
of the fields ${\bn}\equiv\nu^c$ is read off from the
decomposition given above \cite{ag2}:

 \be  M^{\bn}_{AB}= -4i\sq
f'_{AB}<\Sigb^{(R+)}_{44}> =4\sq f'_{AB}\ssb   \ee

and is of the order of the Unification scale  or  somewhat less.
 The left handed or SM neutrinos  receive a direct
Majorana mass from the so called Type II seesaw
mechanism\cite{seesaw2} when the left handed triplet $\bar O$
contained in the ${\bf{\oot}}$ field obtains a vev.
 One obtains\cite{ag1} :

    \be
 M^{\nu}_{AB}= 4\sq f'_{AB}<{\bar O}^{11}> =
 8i f'_{AB}<{\bar O}_{-}>
\ee

The vev   $<{\bar O}_{-}> $ :    $(\ovt,3_L,1)_{\Sigb}$
   arises from a tadpole following   $  SU(2)_L$ breaking (see below).

 The final component of the Seesaw is the Neutrino Dirac mass
 which links the left and right handed
neutrinos :

 Dirac mass :

 \be
 m^{\bn D}_{AB} = 2 \sq h'_{AB} <h^{(1)}_{ 2}> + 4 i {\sqrt 6}
f'_{AB} <h^{(2)}_{ 2}>
\ee

  \vspace{.7 cm}

To determine the Majorana mass terms of the left handed neutrinos
in the effective MSSM we must eliminate the superheavy  $\bn $ and
evaluate the Type II Seesaw tadpole. One then obtains\cite{ag2}
(some factors of $\sqrt{2} $ have been corrected  relative to
eqns(77-79) of\cite{ag2}). \be M_{\nu}^I = -\frac{1}{4} m_{\nu}^D
M_{\bar\nu}^{-1} m_{\nu}^D =
 -r_4 (\hat{h}- 3\hat{f})\hat{f}^{-1}(\hat{h}-3\hat{f})\equiv -r_4 {\hat n}\nnu\\
 \ee
\be
 M_{\nu}^{II} = 8if_{AB}^{\prime}  < {\bar O}_{-}>= r_3 \hat{f}
  \ee
\bea
   r_3 &=&-2 i {\sqrt 3} (\alpha_1 \gamma + 2 {\sqrt 3} \eta \alpha_2)
 ({{\alpha_4}\over {\alpha_2 }}) ({v \over {M_O}}) Sin\beta\nnu\\
 r_4 &=&{{-i\alpha_2 Sin\beta  }\over {4 {\sqrt 3}  }} {{  v }\over {\bar \sigma }} \\
 M_O &=&  2 (M + \eta (3a-p))  )\label{gutseesaw}\nnu\eea
Thus we see that the fermion mass formulae are completely determined
 in terms of the GUT scale breaking parameters($\xi,\lambda,\eta,\gamma,{\bar\gamma},m $),
 the $10 + \oot$ Yukawa couplings(15 parameters)
and the low energy parameters $v_{EW}=174 GeV, tan \beta$.

To perform the fit, we must match the fermion masses and mixings
of the MSSM   RG-extrapolated to the GUT scale $M_X$  :
\bea
L^{FM}_{MSSM}(M_X)   &=&  l^{c^T} D_l l + u^{c^T} D_u u + d^{c^T}
 D_d d \nnu \\
 &+&
  \bar l \not W \nu + \bar U \not Wd' + \nu^T
  ({\cal P} D_\nu {\cal P}^T)\nu + \cdots    \eea

\noindent with the effective theory derived from the MSGUT by
integrating out the heavy fields at $M_X$. Here the
$D_{l,u,d,\nu}$,  C, $ {\cal P} : $ are the diagonal fermion mass
matrices (with mass eigenvalue components $l_A, u_A, d_A, \nu_A
$),  the CKM  mixing matrix and
 the PMNS matrix (Majorana Neutrino mixing)
at the scale $M_X$ in some {\it{fixed}} phase convention for the
MSSM masses and mixings (either at $M_Z$ or at  $M_X$) and
$d'=Cd$.   For example the convention that all the diagonal masses
are real and positive. Currently each group of FM fitters uses
idiosyncratic phase conventions. A  standard format  presentation
of the data of the MSSM  at $M_X$ is  needed and is being pursued.
The corresponding quadratic terms in the effective theory derived
from the GUT are (GUT fields carry carets) :

\bea L^{(2)}_{GUT} &=& \widehat{u^c}^T (\widehat{h} + \widehat{f})\widehat{u}
  + \widehat{d}^c (\widehat{h} r_1 + \widehat{f} r_2)\widehat{d} +
  \widehat{l^c}^T(\widehat{h}r_1 - 3r_2 \widehat{f})\widehat{l}\\
 & + & \widehat{\nu}^T \widehat{M}_{\nu} \widehat{\nu}    +
  \bar {\hat l} \not W {\hat \nu} + \bar {\hat  u} \not W  {\hat d}  \nonumber\eea

where

 \be{\hat M_{\nu}}= r_3 {\hat f}  - r_4 (\widehat{h}+ \widehat{f})\widehat{f}^{-1}
 (\widehat{h}+ \widehat{f})\equiv r_3 {\hat f}  - r_4  \widehat{n} \ee

When equating the two quadratic forms we must allow for Unitary
transformations between the fields in the two Lagrangians :

\bea  L^{(2)}_{GUT} &=& L^{FM}_{MSSM}(M_X) \nonumber\\
  \nonumber\\
     \left( \begin{array}{c} u\nonumber\\
  \nonumber\\
  d' \end{array} \right) &=&  {\cal Q} \left( \begin{array}{c} \widehat{u} \\
  \nonumber\\
  \widehat{d} \end{array} \right); \qquad \qquad  \left( \begin{array}{c} \nu\nonumber\\
  \nonumber\\
  l \end{array} \right) = {\cal L}  \left( \begin{array}{c} \widehat{\nu} \\
  \nonumber\\
  \widehat{l} \end{array} \right)  \\
  \nonumber\\
  u^c &=& {\bf  V}_{(u^c)}\widehat{u^c},\qquad  d^c = {\bf  V}_{(d^c)}
  \widehat{d^c}\qquad  l^c = {\bf  V}_{(l^c)}\widehat{l^c}.
  \nonumber\eea

The unitary matrices ${\bf {\cal Q},{\cal L},{ V}_{(u^c)},{
V}_{(d^c)},{  V}_{(l^c)}}$ describe the
 embedding of the extrapolated MSSM within the MSGUT.  The ($d=5,6$) effective lagrangian
( ${\cal L}^{\Delta B\neq 0}_{eff}(\hat \psi)$) must be
transformed to the extrapolated MSSM basis  in order to derive
rates for such  exotic processes. {\it{Thus  it is clear that
these matrices are neither unphysical nor conventional once the
conventions of the MSSM parameters are fixed}}. In fact we   argue
that {\it{the crucial information given to us by the fitting
procedure is not a prediction of the neutrino masses and mixings
but rather information on these embedding matrices}}.

Since the ${\bf{10,\oot }}$ Yukawas are symmetric it follows that
only two (say ${\bf {\cal Q},{\cal L}}$ ) of these matrices are
independent while the others (${\bf V}_{(u^c)},{\bf
V}_{(d^c)},{\bf  V}_{(l^c)}$) can be determined in terms of the
two chosen to be independent {\it{plus arbitrary diagonal unitary
matrices}} $\Phi_u, \Phi_d, \Phi_l$. Once this is done we obtain :

\bea
 \Phi_u^*  {\bf  V}_{(u^c)}&=&  C \Phi_d^* {\bf  V}_{(d^c)}={\cal Q} \qquad\qquad  \Phi_l^*
{\bf  V}_{(l^c)} ={\cal L} \nonumber\\
 \nonumber\\ \widehat{h}+ \widehat{f} &=& {\bf  V}_{(u^c)}^T D_u {\cal Q} =
 {\cal Q}^T D_u' {\cal Q} \equiv {\cal Q}^T  \Phi_u D_u {\cal Q} \nonumber\\
 \nnu\\  \widehat{h} r_1 + r_2 \widehat{f} & =& {\bf  V}_{d^c}^T D_d C^{\dagger}  {\cal Q}
  = {\cal R}^T D_d'{\cal R}\equiv {\cal R}^T  \Phi_d D_d {\cal R}  \nonumber\\
 \\
  \widehat{h} r_1 -3 r_2 \widehat{f} & =& {\bf  V}_{l^c}^T D_l {\cal L} = {\cal L}^T D_l'
  {\cal L} \equiv {\cal L}^T  \Phi_l D_l
  {\cal L} \nonumber\\
  \nonumber\\
  r_3 {\hat f}  - r_4  {\widehat{n}}  &=& {\cal L}^T {\cal P} D_{\nu} {\cal P}^T {\cal L}
\label{r3r4}\nnu\eea

where we have defined  ${\cal R}=C^{\dagger} {\cal Q}$.  The phase
freedoms $\Phi_u, \Phi_d $ have been found\cite{gohmoh,babmacesnu}
to be important for arranging the tunings that underlie the
successful Type I and Type II fits of the fermion masses. However
the phases $\Phi_l$ play no
 important role so far since the phases in the PMNS matrix $\cal{P}$ are
unknown at present.  If we reabsorb the  phases in  the equation from ${\hat M}_l$   in $ \cal{L}$ then
we also need to redefine the PMNS matrix $\cal{P}$ to reabsorb them  in the neutrino mass equation
(last of  equations  (\ref{r3r4})). This ambiguity should be kept in mind when deducing
predictions of Leptonic  phases from the fit, but does not play any role at this stage.

 We remind the reader that all the parameters   $r_i ; i=1,2,3,4$ are known
in terms of the  MSGUT  parameters but the explicit form is not
used or invoked when solving the fitting
problem\cite{gohmoh,bert}. Rather the parameters $r_i$ and the
Yukawa couplings are determined in terms of the extrapolated
experimental data. Thus the compatibility of the FM fits with the
MSGUT remains to be verified for each prima facie viable fit since
the ability of the MSGUT to reach the required values of the $r_i$
simultaneously while preserving the constraints  of  perturbative
gauge unification  is not obvious.

\subsection{Solution of the fitting problem}

The equations for the down fermion and charged lepton masses may be immediately
solved to yield

\bea
 \widehat{h} r_1 &=& \frac{1}{4} ({\cal L}^T D_l {\cal L} + 3{\cal R}^T D_d' {\cal R})\nonumber\\
 \\
  \widehat{f} r_2 &=& \frac{1}{4} (-{\cal L}^T D_l {\cal L} + {\cal R}^T D_d' {\cal R})
  \nonumber \eea

we define

\be  {\cal D} = {\cal R}^{\ast} {\cal L}^T,\qquad \qquad   {\check
u} =C^T D_u' C   \ee

for later convenience.

  Consider first the case\cite{gohmoh,bert} where the mixing matrices, (hatted)
Yukawa couplings and parameters are all assumed real.  Moreover
the phases $\Phi_u, \Phi_d$  are simply signs  and the CKM phase
$\delta $ is also a sign which in fact is found \cite{gohmoh,bert}
to be minus in ``successful'' fits.
  Then solving the Up quark equations    and eliminating  $r_1,r_2$  using 23 component
and trace we can put it in the form

\be   X\equiv X_l\equiv  {\cal D}   D_{l} {\cal D}^T =  {X_{23}\over  {\check u}_{23} }
 ( {\check u}  -{T_u'\over{T_d'}} D_d') +{T_l\over T_d'} D_d' \ee

Here $  T_f '= tr[D_f']  $. Notice that the non diagonality of
$X_l$ is completely driven by the non-diagonality of the matrix
$u'$ which in turn follows from that of the CKM matrix.
Following\cite{gohmoh} it is convenient to rescale each of the
diagonal fermion mass matrices by the mass of the third generation
fermion of that type to get a dimensionless (tilde-ed) form of the
equations :

 \be D_f \equiv m_{f_3} D_{\tilde f} \qquad  \ee

\be   {\tilde X}_{ l}\equiv {\cal D} D_{{\tilde l}} {\cal D}^T =
  {{\tilde X}_{23}\over {\tilde {\check u}}_{23}} ({\tilde {\check u}}
-{T_{\tilde u'}\over{T_{\tilde d'}}} D_{\tilde d'}) + {T_{\tilde l}\over T_{\tilde d'}} D_{\tilde d'} \ee

The matrix  $X_{\tilde l}$ has  eigenvalues   ${\tilde l}_{1,2} $ and $ {\tilde l}_3=1   $
 by construction and hence it follows that we must have

 \be  det (X_{\tilde l} -{\tilde l}_i I_3) = 0  \ee

Thus we obtain three coupled non-linear equations for ${\tilde
X}_{23}$ and two other quantities which are conveniently chosen to
be  ${\tilde d}_{1,2}$.
   We can solve numerically for   $ {\tilde X}_{23} $ and $ {\tilde d}_{1,2} $,
given any set of up quark and charged lepton masses together with CKM data. Obviously only solutions
within the error bars(usually allowed  to be 1-$\sigma$) for the down quark masses are accepted.
Note however that inasmuch as there is a strong inter-generational hierarchy
for the charged fermion masses   the numerical solution of the three coupled non-linear equations is an
extremely delicate operation requiring utmost care and diligence to find the correct solutions.
Since brute shooting  usually fails in multidimensional problems of this delicacy
approximate analytic  solutions obtained by expanding in the light fermion masses are used to guide the
numerical search for solutions.

  Assuming this has been done  the matrix  $X_l$  is  completely determined
 so that on diagonalizing it one obtains the
crucial matrix $    {\cal D} = {\cal R}  {\cal L}^T $. With ${\cal D} $ in hand
one can proceed by choosing the convenient $\hat f$-diagonal basis mentioned earlier. Rescaling

 \be {\tilde{\widehat f}}= {\widehat f}/m_t = {\tilde{\widehat r_2}} {\cal L}^T X_{\tilde f}
  {\cal L}  \ee

 where we have defined $X_{\tilde f}$ and unitary ${\cal F}  $ by

 \be X_{\tilde f} ={\cal D}^T D_{{\tilde d}}' {\cal D} -D_{\tilde l} ({m_{\tau}\over m_b})  \equiv
  {\cal F} D_{\tilde f} {\cal F}^T    \ee

Since $\hat f$ is diagonal by choice of basis  it immediately follows that

 \be {\cal L} = {\cal F} \Rightarrow     {\cal R}={\cal D}{\cal L} \ee

  Now  since ${\cal L},{\cal R}$
are known and since   ${\tilde{\widehat r}}_{1,2} ={m_b \over m_t}
{\widehat r}_{1,2} $   are calculable   in terms of   $X_{\tilde l}$ it is clear that
 we  have also determined  the Yukawa coupling $\hat h$  :

 \bea {\tilde{\widehat h}}= {\widehat h}/m_t = {\tilde{\widehat r_1}} {\cal L}^T
[3 {\cal D}^T D_{{\tilde d}}' {\cal D} + D_{\tilde l} ({m_{\tau}\over m_b})   ] {\cal L}  \eea

In other words one finds that the fitting of charged fermion mass ratios  requires tuning
of the down quark mass ratios ${\tilde d}_{1,2}$ to less than  one part in $10^{-3}$
for given precise values of up quark and charged lepton masses together with CKM data and yields
the  dimensionless matrices ${\tilde{\widehat h}},{\tilde{\widehat f}}$ and the
exotic embedding matrices ${\cal L},{\cal R}$  and,  given $m_t$,    ${ {\widehat h}},{ {\widehat f}}$.

In the realistic case when the  parameters are complex  a similar but numerically even more difficult
procedure is followed.  Like the sign freedom  in the real case the phase freedom of
choosing $\Phi_u, \Phi_d $  is found
\cite{gohmoh,bert,babmacesnu} to be crucial to obtaining a successful fit.
  The   u equation   still has the same form

 \be   X_l\equiv  {\cal D}   D_{l} {\cal D}^T = p {\check u}  + q D_d        \ee

but now  $Tr X_l\neq Tr D_l$.

 To proceed we solve the    23, 33 components to eliminate $p,q$ and get

 \bea  {\tilde X}_{ l}\equiv {\cal D} D_{{\tilde l}} {\cal D}^T &=&
{{\tilde X}_{23}\over {\tilde  {\check u}}_{23}} {\tilde {\check
u} }   +
  ({ {\tilde X}_{33}- {{\tilde X}_{23}\over {\tilde {\check u}}_{23}}
  {\tilde  {\check u} }_{33}} )D_{\tilde d}
    \label{Xlc} \eea

 Then to determine ${\cal D}$  we must diagonalize

 \be X_{\tilde l} X_{\tilde l}^{\dagger}  \equiv {\cal D} D_{{\tilde l}}^2 {\cal D}^{\dagger}   \ee

 This matrix has
 eigenvalues ${{\tilde l_1^2},{\tilde l_2^2},1}$ and an inspection of its explicit form shows that
it  requires knowledge of  $|X_{23}|, |X_{33}|,\phi_X= Arg( X_{23})-Arg(X_{33})$
 {\it and}  a choice\cite{gohmoh, bert, babmacesnu} of the   quark
mass phases $\Phi_u, \Phi_d $.  The numerical results obtained by
these authors are then a specification of the right hand side of
eqn(\ref{Xlc}) consistent with some acceptable values of the
charged lepton masses. We refer the reader to the original papers
for the procedure for fixing  the unknown phases. Once this rather
horrendous numerical problem has been solved (its trickiness
accounts for the fact that even 12 years after it's solution was
first explicitly attempted\cite{babmoh} this system continues to
throw out surprises\cite{babmacesnu}) one can proceed essentially
as before :
 diagonalizing the rhs  of eqn(\ref{Xlc})   numerically thus yields the matrix $\cal D$.

  Since ${\widehat f}$  is diagonal and real by convention  one writes

 \be {\tilde{\widehat f}}= \tilde{\widehat f}/m_t = {\tilde{\widehat r_2}}
{\cal L}^T X_{\tilde f}   {\cal L}  \ee

where

 \be
 X_{\tilde f} ={\cal D}^{\dagger} D_{{\tilde d}'} {\cal D}^* -D_{\tilde l} ({m_{\tau}\over m_b})  \equiv
 {\cal F} D_{\tilde f} {\cal F}^T \ee

Since   $X_{\tilde f}$ is   symmetric ,    $   {\cal F} $  is
Unitary
   and  $  D_{\tilde f}  $ is real like ${\tilde{\widehat f}}$, it follows that

 \be  {\tilde{\widehat f}}= |{\tilde{\widehat r_2}} | D_{\tilde f}  \ee

so that

\be{\cal L} ={\cal F}^* e^{{{-i Arg({\tilde{\widehat r_2}})}\over 2}} \ee

 where ${\cal F}$  is  found by diagonalizing
 $ X_{\tilde f} X_{\tilde f}^{\dagger}={\cal F} D_{\tilde f}^2 {\cal F}^{\dagger}$

   Finally
 \be {\cal R} = {\cal D}^* {\cal L}\qquad ;\qquad {\cal Q}={ C}{\cal R}  \ee

  In summary :  for given up and charged lepton masses and given CKM mixing angles, by tuning
${\tilde d}_{1,2},\delta_{CKM}$ within their allowed ranges and for a certain choice of  the phases
$\Phi_u,\Phi_d$, one completely fits
the charged fermion masses and determines the FM Yukawa couplings $h',f'$  of the GUT along
with the exotic embedding matrices  ${\cal L},{\cal Q}, \Phi_u,\Phi_d$.

\subsection{Fitting Neutrino Masses}

The current FM Fitting Furore\cite{gohmoh,bert,babmacesnu}
 was triggered by the remarkably simple observation of\cite{bsv} regarding
naturalness of large atmospheric mixing angles in   MSGUTs with dominant Type II seesaw mechanisms
given the near equality of the  $\tau$ lepton and bottom quark masses at $M_X$. If one assumes Type II
domination i.e $r_3>>r_4$ the Seesaw formula simplifies to
just

\be   {\tilde{\widehat{f}}}   =
({1\over {m_t r_3}}) {\cal L}^T {\cal P} D_{\nu} {\cal P}^T {\cal L}  \ee

Since we chose a$\hat f$-diagonal basis it immediately follows that

 \be {\cal P}={\cal L}^*  e^{{i arg(r_3)}\over 2}   \qquad ;\qquad    D_{\nu} =   |r_3| {\widehat{f}}  \ee

Thus the neutrino mixing angles and ratio of mass squared splittings can be
determined under these assumptions since we know ${\widehat{f}}, \cal{L} $.

In the general case the neutrino mass fitting equations take the scaled form :

\be {\cal N}{ D_{\tilde  N}} {\cal N}^T \equiv  {\tilde{\widehat{f}}} -({r_4 \over {r_3}})  {\tilde n} =
({1\over {m_t r_3}}) {\cal L}^T {\cal P} D_{\nu} {\cal P}^T {\cal L}  \ee

where

 \be {\tilde{n}}= {  n}/m_t =   ( {\tilde{\widehat{h}}}-3 {\tilde{\widehat{f}}} )
 {\tilde{\widehat{f}}}^{-1} ({\tilde{\widehat{h}}} -3  {\tilde{\widehat{f}}})   \ee

Thus the mixing matrix and neutrino masses are also completely determined :

 \be {\cal P}={\cal L}^* {\cal N} e^{{{i Arg({ {  r_3}})}\over 2}}  \qquad ;
\qquad    D_{\nu} = m_t |r_3| {  D_{\tilde N}} \ee

 These fitting problems have been formulated and solved with increasing refinement
by a number of authors\cite{gohmoh,bert,babmacesnu}. It
seems\cite{babmacesnu} that Type II dominant as well as  Type I
and Type I plus Type II combined solutions (which are not
perturbations of Type II dominant solutions ) can be found in the
complex case. Further solutions probably still remain to be found
and the possible solutions definitely still need to be compactly
characterized and parameterized. The fits achieved so far already
motivate a detailed examination of what type of Seesaw is actually
allowed  by the MSGUT in various regions of its parameter space.
We have argued that in each case the exotic embedding matrices are
-like the  `philosophers stone' - the so far unregarded true
product of the fitting calculation.
 We emphasize  that in practice one accepts solutions which give $\nu$ oscillation
parameters which  lie within the $1-\sigma$ ranges around the central values of the
fermion mass and mixing parameters  so that {\it{the true output
of the fitting calculation are the previously completely unknown embedding matrices}}
${\cal Q},{\cal L},\Phi_u,\Phi_d$  {\it{which specify how an MSSM($M_X$) with given
conventions lies within the MSGUT}}. These matrices are crucial for pinning
down the prediction of $\Delta B\neq 0$  processes in the MSGUT.
Before considering that aspect however we turn to the question of
what kinds of solutions are compatible with the MSGUT.

\section{Scanning the MSGUT for Neutrino Masses  and Mixings }

The possibility of large PMNS mixing angles is  well understood in
the Type II dominant case \cite{bsv,gohmoh}, where it appears as a
natural corollary of the approximate unification of the running
bottom quark and tau lepton masses at scales $O(M_X)$ due to the 3
fold faster evolution of the bottom quark mass. As is evident from
eqn.(\ref{r3r4}) the parameter which controls  the strength of
Type I versus Type II Seesaw in the MSGUT is the ratio of the
coefficients $r_4$ and $ r_3$. Complete Type II dominance requires
$r_4 << r_3$.  To illustrate how the MSGUT yields information on
this ratio we work with an example of a quasi-realistic
{\it{real}}  Type II fit that ignores the CP violating phase,
which was kindly provided to us by S.Bertolini and M. Malinsky.
Although Type II and only semi-realistic  it should be emphasized
that the values  of $\hat h, \hat f$ given will be rather typical
since they are fixed by the charged fermion mass data  and further
selection is  then imposed based
 on   compatibility with the neutrino data.

The data of the example solution  ( at $M_X$ and for $ \tan \beta =10$  ) is
\bea
D_u &=&\{0.785556, -191.546, 70000\} MeV \nnu\\
D_l &=& \{0.3585, 75.7434, 1290.8\} MeV\nnu\\
d_3 &=& m_b=1138.07 MeV\nnu\\
\{Sin\theta^c_{12}, Sin\theta^c_{23}, Sin\theta^c_{13}\}&=&
\{0.2229, 0.03652, -0.00319 \}\eea

Notice the negative sign of $u_3$ and $\theta^c_{13}$ (this
corresponds to taking $\delta_{CKM}=\pi$), moreover the fitting
procedure  gives

\bea {\tilde x}_{23} &=& -0.14325 \quad;\quad {\tilde d_1}=
-.001105 \quad;\quad   {\tilde d_2}=-.02747 \eea

so that the sign ambiguity $\Phi_d$ is also fixed.

\be
{\cal D}=  \left( \matrix{ 0.98953 & -0.136941 & 0.045582 \cr -0.128508 & -0.979735 & -0.153642 \cr
 -0.0656981 & -0.146176 & 0.987075 \cr  } \right)
\ee

Supplying  $m_b= 1138.07 \quad  MeV $  allows one to calculate
\be
{\cal L}={\cal F}= \left( \matrix{ 0.794035 & -0.582472 & 0.173881 \cr -0.563923 & -0.599057 &
 0.568437 \cr 0.226934 & 0.549415 & 0.804142 \cr  }\right)
\ee

Then  the matrices $ {\cal R},{\cal Q}$ follow :

\bea
{\cal R}&=&  \left(\matrix{ 0.87329 & -0.469294 & 0.130873 \cr 0.415588 & 0.577356 & -0.702813 \cr
 0.254266 & 0.668149 & 0.699233 \cr  }\right) \nnu\\
{\cal Q}&=& \left(\matrix{ 0.943138 & -0.330924 & -0.0313091 \cr 0.219729 & 0.691351 & -0.688297 \cr
0.24942 & 0.642279 & 0.724753 \cr  }\right) \eea

The scaled Yukawa couplings are \bea
{\tilde{\hat f}}&=&  \left({\begin{array}{ccc} -0.000595629 & 0 & 0 \\
0& 0.0020326& 0 \\
 0&  0 & -0.00804198\\
  \end{array}}\right) \nonumber \\
{\tilde{\hat h}}&=&  \left({\begin{array}{ccc} 0.0626836& 0.159778& 0.181181  \\
 0.159778& 0.409183& 0.466796\\
 0.181181&
    0.466796& 0.532013\\
  \end{array}}\right) \eea

The matrix  ${\hat n}$   multiplying $-r_4$ in the Type I mass is
\bea
{\tilde{\hat n}}&=&  \left({\begin{array}{ccc} 1.4996& 3.8747& 4.55331  \\
 3.8747& 9.98038& 11.6875 \\
 4.55331& 11.6875&
    13.63\\
  \end{array}}\right)
\eea

From these the Type  II mixing angles and the ratio of the 23 and 12 sector mass squared
splittings is found to be
\bea
 \{Sin^2 (2\theta^P_{12}), Sin^2(2\theta^P_{23}), Sin\theta^P_{13}\}
&=& \{0.90981, 0.88873, 0.17387\} \nnu\\
 \Delta m^2_{12}/ \Delta m^2_{23}&=&  0.06237  \eea

The Yukawa couplings obtained  are typical of Type II fits.  It is
clear that unless $r_4/r_3$  is very small the Type I term will
dominate completely. To see just how small $r_4$ should be we plot
the   neutrino mixing angles $ Sin^2 2\theta_{12}, Sin^2
2\theta_{23}$   versus this ratio as Fig 8.

From Fig. 8 it is evident that a value of $|r_4/r_3|>.0003$ causes
a collapse of the large mixing angles of the Type II dominant
solution.  Thus we should not expect a  Type II solution to work
in the MSGUT unless   $R=r_4/r_3$  (which is completely determined
by the GUT   as given in the previous section ) obeys $|R|<
<10^{-3}$. The MSGUT formulae for $r_3,r_4$ are  given in
eqn(\ref{gutseesaw})

 Using these   we  plot  the ratio R versus $\xi$  and  find
 that its typical value is $\sim 10^{-1}$ or more,
not $10^{-3}$ or less. Illustrative plots for the real and complex solutions
of the cubic eqn.(\ref{cubic}), for real  $\xi$  and  $\lambda =.7$ are shown as Fig 9.

 From the plots it is evident that  in the real case R has a chance of being
 of the required small size only in the window near $\xi=-.7$ and
between about $\xi=1 $ and $\xi=4$. However in the former region
one finds that the corrections to $Sin^2\theta_W$ diverge, while
the region $\xi\in (1,4) $ is not allowed by the requirement  that
$\Delta Log_{10} M_X > -1 $. On the  other hand, in the complex
case, the ratio seems bounded below by 1 !  {\it{Thus we see that
if Type II fits were the only allowed ones then the combination of
the gauge unification requirements with those imposed by neutrino
mass phenomenology  tend to rule out the MSGUT based on this class
of solutions}}.
  This conclusion however still requires a more thorough study of all possibilities.
Moreover recent work \cite{babmacesnu} has shown that in fact
large mixing angles {\it{can}} be achieved even in  Type I
solutions and Type I-II combined solutions (which are far from
pure Type II solutions). Our intent, in this talk, is not to
provide an exhaustive survey of the possibilities but only to
illustrate how the combined requirements of neutrino oscillations
and baryon stability can severely constrain the MSGUT over its
parameter space.  A detailed survey of the
  MSGUT for each of the three solutions of the basic
cubic equation to find   in which  regions, if any,
the ratio R matches that required by the FM fit is quite feasible
using our methods, and will be reported elsewhere\cite{newtas}.

Note that while the FM fits described in the previous section do not determine
the over all mass scale of the neutrinos since the input data does not contain
this information, the same is not true  in the context of the MSGUT.  Using the
 dimensionless versions ${\tilde {\hat f}}$ and ${\tilde{\hat h}}$
given by the real Type II FM fitting analysis as a guide to
typical values of the Yukawa couplings allows one to compute  the
magnitudes of all three neutrino masses for each type of fit. When
this is done another problem becomes apparent : except possibly
for narrow ranges of $\xi $ the largest neutrino mass (Type I or
II) is much smaller than the mass splitting known from atmospheric
neutrino oscillations.  Thus even if one can find a region where
the ratio of mass splittings  and mixing angles for neutrinos are
in the allowed region the additional consistency constraint that
$|( m_{\nu} )_{max}| > |(\Delta m )_{atmos}| $ alone can exclude
most of the  parameter space !   An illustrative plot  of  the
maximum type I and Type II neutrino masses for the Yukawa coupling
matrices that arises in the Bertolini and Malinsky solution used
by us for illustration is given for the real solution as Fig. 10.
Due to the cut at $\xi=8.6$
 imposed by the  condition $\Delta Log_{10} M_X > -1 $ we see that there is no region with large enough
values of  $|(m_{\nu} )_{max}|   $.  Similarly in the complex case
we get the plots shown in Fig. 11. As expected the Type I
dominates completely. However the mass magnitudes   for Type I
 tend to be  too small : as can be seen from Fig. 11 and the magnifications shown in Fig 12.   Since the shortfall is
only a factor of 10 and we have not yet used the  complex Type I
fits found in\cite{babmacesnu} it would be premature to rule out
Type I fits.  Nevertheless a certain tension is apparent.

If the impossibility of large enough overall neutrino mass scale
is borne out by a comprehensive analysis\cite{newtas}   it  would
require a revision of the perturbative MSGUT. Similar arguments
were used in\cite{gohmoh04} (for the Type II FM fit case only)  to
motivate an extension of the model by introducing an addition
${\bf{54}}$ plet so as  raise the value of the ${\bf{\oot}}$ vev
to a high scale to allow Type II to dominate {\it{and}} yield a
large enough neutrino mass. Note however that in that case the
minimality of the MSGUT is seriously diluted necessitating a
complete  reanalysis which has not been performed so far. An
alternative to this extension may be to implement a non
perturbative mechanism\cite{tas,newtas} based on   dynamical GUT
symmetry breaking  down to $SU(5)\times U(1) $ at $\Lambda_X\sim
10^{17} GeV$  and then use a Type I fit with small ${\bf{\oot}}$
vev $\bar\sigma$   to achieve a larger value of the overall
neutrino mass scale.

 \section{$\Delta B\neq 0 $ IN MSGUT }

Finally we briefly indicate   the $d=5 $ baryon decay operators  determined by the FM fit.
If any FM fit   is found to  be consistent with  all the constraints discussed
above then the computation of the actual Baryon Decay predictions
 will become a worthwhile exercise.

The basis for this is the effective superpotential that arises when the superheavy Higgs
 triplets are
integrated out from the theory\cite{ag1,ag2} \bea  W_{eff}^{\Delta
B\neq  0 } &=& {\hat L}_{ABCD} ({1\over 2}\epsilon {\hat Q}_A
{\hat Q}_B {\hat Q}_C {\hat L}_D) +{\hat R}_{ABCD}  (\epsilon
{\hat {\bar e}}_A {\hat {\bar f u}}_B {\hat {\bar u}}_C {\hat
{\bar d}}_D) \eea \bea {\hat L}_{ABCD} &=&  {\cal S}_1^{~1} h_{AB}
h_{CD} + {\cal S}_1^{~2} h_{AB} f_{CD} +
 {\cal S}_2^{~1}  f_{AB} h_{CD} + {\cal S}_2^{~2}  f_{AB} f_{CD}  \eea
  \bea {\hat R}_{ABCD} &=& {\cal S}_1^{~1}  h_{AB} h_{CD}
 - {\cal S}_1^{~2}  h_{AB} f_{CD} -
 {\cal S}_2^{~1}  f_{AB} h_{CD} + {\cal S}_2^{~2}  f_{AB} f_{CD} \nonumber \\
 &-& i{\sqrt 2} {\cal S}_1 ^{~4} f_{AB} h_{CD}
+i {\sqrt 2} {\cal S}_2 ^{~4} f_{AB} f_{CD}  \eea
Here   ${\cal S}= {\cal T}^{-1} \quad$ where  ${\cal T}  $  is the  mass matrix
 in the  $[3,1,\pm 2/3]$-sector  which is the representation type that
mediates $d=5$ baryon decay. The Yukawa coefficients $h_{AB},
f_{AB}$ are related to those in the superpotential by   $h_{AB} =
2 {\sqrt 2} h_{AB}', f_{AB} = 4 {\sqrt 2} f_{AB}'    $

 Substituting for the GUT superfields (with carets) in terms of the embedding matrices
${\cal Q},{\cal R}, {\cal L}$ determined by the FM fit

\bea {\widehat Q}_{L} &=& {\cal Q}^{\dagger} {Q}_{L} \qquad\qquad
{\widehat L}_{L} = {\cal L}^{\dagger} {L}_{L} \\
{\hat {\bar u}}={\hat u_c}  &=& {\cal Q}^{\dagger} \Phi_u^* u_c
\qquad\qquad {\hat {\bar d}}={\hat d_c}  = {\cal  R}^{\dagger}
\Phi_d^*{d_c} \qquad {\hat {\bar l}}= {\hat l_c} = {\cal
L}^{\dagger} \Phi_l^*{l_c} \nnu\eea

one obtains the coefficients of mass eigenstate MSSM fields  in
the $\Delta B\neq 0$ superpotential to be \bea { L}_{ABCD} &=&
{\cal Q}^{\dagger}_{AE} {\cal Q}^{\dagger}_{BF} {\cal
Q}^{\dagger}_{CG}
 {\cal L}^{\dagger}_{DH} {\hat L}_{EFGH}\nnu\\
{ R}_{ABCD} &=& {\cal L}^{\dagger}_{AE} {\cal Q}^{\dagger}_{BF}
{\cal Q}^{\dagger}_{CG} ({\cal R}^{\dagger})_{DH} (\Phi_l^*)_{EE}
(\Phi_u^*)_{FF} ( \Phi_u^*)_{GG}  ( \Phi_d^*)_{HH}{\hat R}_{EFGH}
\eea

A  similar calculation\cite{newtas} can be done to determine the
effective operators for $d=6$ baryon violating operators, which
will be relevant even if the Supersymmetry breaking scale
eventually turns out to be large enough ($> 100 $TeV) to exclude
any observable $d=5$ baryon decay. Similar care needs to be
exercised when studying lepton flavour violation.

\section{Conclusions and Outlook}

The MSGUT based on the ${\bf{210\oplus126\oplus\oot\oplus 10}}$
Higgs system is the simplest Supersymmetric GUT that elegantly
realizes the classic program of Grand
 Unification and
which is prima facie compatible with all known data. Its symmetry
breaking structure is so simple as to permit an explicit analysis
of its mass spectrum at the GUT scale and an evaluation therefrom
of the threshold corrections  and mixing matrices relevant to
various phenomenologically important quantities. Since it has the
least number of parameters of any theory that accomplishes as much
this theory currently merits the name of the minimal
supersymmetric GUT or MSGUT. The same simplicity and analyzability
of GUT scale structure also applies to the theory with an
additional {\bf{120}}-plet, since it contains no standard model
singlets, and thus justifies calling it the nMSGUT.

  The small number of Yukawa couplings of the MSGUT makes the fit to the now well
 characterized fermion mass spectra  very tight so that the combined
constraints of this fit and the preservation of the gauge unification
observed at one loop in the MSSM  may well be enough to rule out most or
even all of the parameter space of this theory.
A corresponding investigation is also possible for the nMSGUT  and both are
 in progress\cite{newtas}.

We have emphasized the need for clarity regarding the flavour basis used when
performing the FM fit.
When this is maintained it becomes obvious that  the true outputs of the FM fitting
calculation are really the embedding matrices that define how the MSSM,
with fixed phase conventions,  lies within the MSGUT.
With these matrices determined by the FM fit in hand,
  one will be in a position to perform a much more  reliable
calculation of $d=5,6$ Baryon decay in the MSGUT with the Susy
breaking scale as the chief remaining uncertainty. This
realization underlines and emphasizes the organic connection
between the physics of neutrino mass and Baryon
violation\cite{babpatwil} i.e between the physics of ultraheavy
and ultra-light particles which is the most intriguing implication
of the discoveries of Super-Kamiokande.

Our discussion concerning embedding angles also has  implications for a treatment of
Lepton Flavour violation in the MSGUT which will be worth exploring in detail
 if  a viable region of the parameter space of the perturbative MGUT emerges.
  On the other hand if no such candidate region is available then the remaining
options that will need to be explored could  consist on the one hand of the
analogous calculations in the nMSGUT, namely when an additional 120-plet is introduced :
 which  radically enlarges the possibilities as far as FM fitting
is concerned\cite{bert}.  Or the MSGUT could be extended by
 engineering the models to ensure  Type II
 dominance\cite{gohmoh04}.

 Another  possibility  is that the
symmetry breaking at the GUT scale is primarily determined not
by the perturbative superpotential
but rather by a non-perturbative mechanism\cite{trmin,tas} whereby gaugino
condensation in the
coset SO(10)/H (where H could be, e.g, $SU(5)\times U(1)$ or $ SU(5)$) drives a
H-singlet  Chiral condensate of (say) the 210-plet field at a scale
$\sim \Lambda/\lambda^{\alpha} > M_X$. In that case the spectra given by us in terms
of the vevs $a,p,\omega$ are still of use but they are no longer determined by the
cubic equation(\ref{cubic}). Rather after breaking the symmetry to the group H
at a scale higher than the perturbative scale $M_X$ one would examine the symmetry
breaking in the effective $SU(5)$ symmetric theory at lower energies $\sim M_X$.
 Such a scenario
 would thus  not only utilize the problematic strong coupling regime lying just above
the perturbative unification scale for dynamical symmetry breaking
of the GUT theory at a scale $\Lambda_X$ ( which is  determined by the low energy data
and the Grand Unified structure and functions as an internally defined
upper cutoff for the MSGUT) but also provide an explanation for the
``SU(5) conspiracy"\cite{abmrs01}
that seems to operate within  SO(10) Susy GUTs when proper account is taken of  neutrino data,
superheavy Susy spectra  and RG evolution.  This kind of scenario  would fit naturally
 with a Type I mechanism which favours small values of the ${\bf{\oot}}$ vev. Thus it contrasts sharply
with the proposal of\cite{gohmoh04} and should be distinguishable
phenomenologically from it.

  We conclude with a tentative proposal  to reconcile String theory based models and
the ${\bf{\oot}}$ based Type I and Type II seesaw mechanisms that
occur in  RGUTs. Recall that String theory, particularly with
level 1 Kac-Moody algebras, does not favour the emergence of
effective  GUTs  containing  adjoint and larger representations as
its massless sector\cite{stringuts}. Even the use of higher KM
levels permits the occurrence of  only very restricted numbers and
types of higher GUT representations. In particular, in the case of
SO(10), one cannot enlist the type of combinations characteristic
of RGUTs (${\bf{45\oplus 54 \oplus \oot \oplus 126}}$  or ${\bf{
210 \oplus \oot \oplus 126}}$) etc. These difficulties  led to a
decline in attempts to SO(10)  GUTs  from string theory. However
of late the growing appreciation of the naturalness of SO(10)
unification in the light of discovered neutrino mass
\cite{csarevs}  has motivated  renewed effort in this
direction\cite{newstringuts}. SO(10) type families are generated
with gauge symmetry pre-broken to MSSM or somewhat larger. However
the issue of implementing the seesaw mechanism whether Type I or
Type II in the way achieved so naturally in RGUTs remains
problematic due to the difficulty of in building models in which
the  ${\bf{ \oot \oplus 126}}$ representations remain  massless in
the string model.

   A way out of this difficulty
may perhaps be found by appreciating that in RGUTs the  ${\bf{ \oot \oplus 126}}$
 fields are superheavy
and their neutral components have either a superlarge vev
(corresponding to $M_{B-L}\sim M_X$ ) or   very small vevs ($\sim
M_W $ or $\sim  m_{\nu}$ ).  The former kind of vev is that of the
   right handed triplets that give rise to the right handed neutrino's
   superlarge Majorana mass and thus
 a small Type I seesaw mass for the MSSM neutrino.    Its large size is
 compatible with the ``pre-broken''
structure of ``string derived GUTs''   where the breaking of the SO(10)/GUT
  gauge symmetry
 still discernible in the matter super multiplet structure  is accomplished
 already at the level
 of defining the light sector of the  String theory and in an effective
 description corresponds
to a  vevs of {\it{superheavy}} fields in the appropriate large
chiral representations (such as ${\bf{45,  54, 210,  \oot,126}}$
etc ).
  On the other hand  the light vev $\sim M_W <M_S\geq 1TeV $  in MSSM doublet channels
occurs only {\it{after}} breaking  of supersymmetry and its small size is then thought to be
naturally compatible with the status of Susy breaking as a tiny correction to the pre-broken
 String GUT picture which is  supersymmetric, conformal etc and  therefore its derivation
 can be legitimately postponed.  Furthermore the Type II seesaw mass generating vev
 of the left handed triplet in the ${\bf{  \oot  }}$ is an even higher order effect that
  arises due to a tadpole
{\it{in a superheavy field}} induced once EW symmetry breaking has
taken place due to the coupling of this superheavy field with the
doublets that get a small EW weak vev.

 Thus it appears that the  ${\bf{  \oot \oplus 126}}$ fields
{\it{ need no longer be sought in the massless sector of the String theory}}. Instead  it is
 sufficient to investigate whether the superheavy
``Higgs Channel''  corresponding to  ${\bf{ \oot  }}$ or the above
mentioned  relevant sub-representations in fact couple to the
putative light fields in an appropriate way.  To put it simply the
implementation of the $R_p$ preserving seesaw mechanism in Stringy
GUTs may   require  a small ``leakage'' connecting the superheavy
``field'' in the ${\bf{  \oot }}$ channel  to the matter fermions
and light doublets and the   availability of the control parameter
 $M_W/M_X$  provides a natural way to keep the destabilizing effects
  of such heavy-light couplings
under control : given that some way has somehow been found to
break Susy   and generate the EW scale in the first place ! This
refinement of the effective theory paradigm used to extract the
low energy theory from string models is both novel and  consonant
with the characteristic and elegantly consistent tying together of
very large and very small mass scales achieved by the seesaw
mechanism.

 \vspace{.5 true cm}
\section{Acknowledgments}
 \vspace{.5 true cm}
I thank Sumit Garg for collaboration and technical  help while
studying   FM fitting in the MSGUT and S. Bertolini and M.Malinsky
for providing me with a complete set of data for one of their FM
fits. It is a pleasure for  to acknowledge stimulating discussions
with K.Babu, Stefano Bertolini, I. Dorsner, Ilya Gogoladze, M.
Malinsky, R.N. Mohapatra, S. Raby, G.~Senjanovi\'c, F.~Vissani,
during the Conference Plank05 at Trieste May 23-28, 2005. I am
grateful to to the   G. Senjanovic and High Energy Group of the
Abdus Salam ICTP, Trieste, K. Huitu and the Theory Group,
Institute of Physics, Helsinki, and and the Theory Group, CERN,
Geneva for hospitality  during the writing of these proceedings
and A. Faraggi and H.P. Nilles for discussions concerning the
possibility of implementing
 R-parity preserving  seesaw and $SO(10) $  in  string theory.

\newpage
 \begin{table}
$$
\begin{array}{l|l|l}
{\rm Field }[SU(3),SU(2),Y] &  PS  \qquad  Fields  & {\rm \qquad Mass}  \\
 \hline
 &&\\
 A[1,1,4],{} \bar A[1,1,-4] &{{\s^{44}_{(R+)}}\over
\sq}, {{\os_{44(R-)}}\over \sq}&
 2( M + \eta (p +3a +  6 \omega )) \\
 C_1[8,2,1],{} \bar C_1 [(8,2,- 1] &
\s_{\bn\alpha\dot1}^{~\bar\lambda},{}
 \Sigb_{\bn\alpha\dot 2}^{~\bar\lambda} &
 2 (-M + \eta (a+\omega)) \\
 &&\\
C_2[8,2,1],{}\bar C_2 [(8,2,- 1] &
\Sigb_{\bn\alpha\dot1}^{~\bar\lambda},{}
 \s_{\bn\alpha\dot 2}^{~\bar\lambda} &
 2 (-M + \eta (a-\omega)) \\
D_1[3,2,{7\over 3}],{}\bar D_1 [(\bar 3,2,- {7\over 3}] &
\Sigb_{\bn\alpha\dot1}^{~4},{}
 \s_{4\alpha\dot 2}^{~\bn} &
 2 (M + \eta (a+\omega)) \\
 &&\\
D_2[3,2,{7\over 3}],{}\bar D_2 [(\bar 3,2,- {7\over 3}]
 & \s_{\bn\alpha\dot1}^{~4},{}
 \Sigb_{4\alpha\dot 2}^{~\bn} &
 2 (M + \eta (a+3\omega)) \\
 &&\\
E_1[3,2,{1\over 3}],{}\bar E_1 [(\bar 3,2,- {1\over 3}] &
\Sigb_{\bn\alpha\dot 2}^{~4},{}
 \s_{4\alpha\dot 1}^{~\bn} &
 - 2 (M + \eta (a - \omega)) \\
K[3,1,-{8\over 3}],{}\bar K [(\bar 3, 1, {8\over 3}] & \Sigb_{\bn
4(R-)},{}
 \s^{\bn4}_{(R+)} &
 2 (M + \eta (a+p+ 2 \omega)) \\
 &&\\
L[6,1,{2\over 3}],{}\bar L [(\bar 6,1, -{2\over 3}] &
  (\Sigb_{\bm\bn}^{'(R0)},
 \s^{'\bm\bn}_{(R0)})_{\bm\leq\bn} &
 2 (M + \eta (p -a)) \\
& \Sigb'_{\bm\bn}=\Sigb_{\bm\bn},{} \bm\neq \bn&\\
&\Sigb'_{\bm\bm}={{\Sigb_{\bm\bm}}\over\sq}&\\
 &&\\
   M[6,1,{8\over 3}],{}{\ovl M} [(\bar 6,1, -{8\over 3}] &
  (\Sigb^{'(R+)}_{\bm\bn(R+)},{}
 \s^{'\bm\bn}_{(R-)})_{\bm\leq\bn} &
 2 (M + \eta (p -a + 2 \omega )) \\
N[6,1,-{4\over 3}],{}\bar N [(\bar 6,1, {4\over 3}] &
 (\Sigb_{\bm\bn}^{'(R-)},
  \s^{'\bm\bn}_{(R+)} )_{\bm\leq\bn} &
 2 (M + \eta (p -a-2\omega )) \\
 O[1,3,-2],{}\bar O [(1,3, +2] &
 {{{\vec\s}_{44(L)}}\over \sq},{}
 {{{{\vec{\Sigb}}^{44}_{(L)}}}\over \sq} &
 2 (M + \eta (3a-p)) \\
 &&\\
 P[3,3,-{2\over 3}],{}\bar P [\bar 3,3, {2\over 3}] &
 {\vec\s}_{\bm 4(L)},
  {\vec\Sigb}^{\bm 4}_{(L)}  &
 2 (M + \eta (a-p)) \\
  &&\\
W[6,3,{2\over 3}],{}{\overline W} [({\bar 6},3, -{2\over 3}] &
 {{{\vec\s'}_{\bm\bn(L)}}},
 {\vec\Sigb}^{\bm \bn}_{(L)}  &
 2 (M - \eta (a+p)) \\
I[3,1,{10\over 3}],{}\bar I [(\bar 3,1,- {10\over 3}] &
\phi_{~\bn(R+)}^4,{}
 \phi_{4(R-)}^{~\bn} &
 -2 (m + \lambda (p+a+4\omega)) \\
S[1,3,0] & \vec\phi^{(15)}_{(L)} & 2(m+\lambda(2a-p))\\
Q[8,3,0]& {\vec\phi}_{\bm(L)}^{~\bn}&
 2 (m - \lambda (a +p)) \\
U[3,3,{4\over 3}],{} \bar U[ \bar 3,3,-{4\over 3}] &
{\vec\phi}_{\bm(L)}^{~4},{} {\vec\phi}_{4(L)}^{~\bm}&
 -2 (m - \lambda (p-a)) \\
 &&\\
V[1,2,-3],{} \bar V[ 1,2,3] & {{{\phi}_{44\alpha\dot
2}}\over\sq},{} {{\phi^{44}_{\alpha\dot 1}}\over \sq}&
 2 (m  + 3 \lambda (a + \omega)) \\
B[6,2,{5\over 3}],{}\bar B [(\bar 6,2, -{5\over 3}] &
 (\phi_{\bm\bn\alpha\dot 1}',
 \phi^{'\bm\bn}_{\alpha\dot 2} )_{\bm\leq\bn} &
 -2 (m + \lambda (\omega -a )) \\
Y[6,2,-{1\over 3}],{}\bar Y [(\bar 6,2, {1\over 3}] &
 (\phi_{\bm\bn\alpha\dot 2}',
\phi^{'\bm\bn}_{\alpha\dot 1})_{\bm\leq\bn} &
 2 (m - \lambda (a+\omega )) \\Z[8,1,2],{} \bar Z[ 8,1,-2] & {\phi}_{~\bm(R+)}^{\bn}
{\phi}_{\bm(R-)}^{~\bn}&
 2 (m + \lambda (p-a)) \\
\end{array}
$$
\label{table I} \caption{{\bf{i)}}     Masses   of the unmixed
states in terms of the superheavy vevs. The $SU(2)_L$ contraction
 order is always $\bar F^{\alpha} F_{\alpha} $. The
 primed fields defined for $SU(3)_c$ sextets
  maintain unit norm. The absolute value
   of the expressions in the column ``Mass" is understood.}
\end{table}
 \newpage

 {\bf {Appendix   : Tables
of masses and mixings }} \vskip.5 true cm
 \vspace{.2 true cm}

In this appendix we collect our results for the chiral fermion/gaugino states,
masses and mixing matrices for the reader's convenience.
   Mixing matrix rows are labelled by barred
  irreps and columns by unbarred. Unmixed cases({\bf{i)}}) are given as Table
  I.

 {\bf{ ii)\hspace{ 1.0 cm} Chiral Mixed states}}\hfil\break

\vspace{.3 cm}

 {\bf{a)}}$ [8,1,0](R_1,R_2)\equiv (\hat\phi_{\bm}^{~\bn},\hat\phi_{\bm
(R0)}^{~\bn})  $

 \be
{\cal{R}} = 2 \left({\begin{array}{cc} (m-\lambda a ) &
-\sqrt{2}\lambda\omega \\ -\sqrt{2}\lambda\omega & m+\lambda( p-a)
\end{array}}\right)
\ee

\be m_{R_{\pm}}=   |{\cal R}_{\pm}| = |2 m [1 +({\pt \over 2} -
\at) \pm \sqrt{({\pt\over 2})^2 + 2 \omt^2}]| \ee
 The  corresponding
eigenvectors can be found by diagonalizing the matrix ${\cal
R}{\cal R}^{\dagger}$. \vspace{.3 cm}

{\bf{b)}}\hspace{ 1.0cm}  $ [1,2,-1]({\bar h}_1,\bh_2,\bh_3,\bh_4)
\oplus     [1,2,1](h_1,h_2,h_3,h_4) $\hfil\break
 $.\qquad\qquad\equiv
(H^{\alpha}_{\dot 2},\Sigb^{(15)\alpha}_{\dot 2},
\s^{(15)\alpha}_{\dot2},{{\phi_{44}^{\dot 2\alpha}} \over
\sq})\oplus (H_{\alpha {\dot 1}},\Sigb^{(15)}_{\alpha \dot1},
\s^{(15)}_{\alpha\dot 1}, {{\phi^{44\dot 1}_{\alpha}} \over \sq})
 $

\bea {\cal{H}}=\left({\begin{array}{cccc} -M_{H} &
+\overline{\gamma}\sqrt{3}(\omega-a) & -{\gamma}\sqrt{3}(\omega+a)
& -{\bar{\gamma}{\bar{\sigma}} }\\
 -\overline{\gamma}\sqrt{3}(\omega+a) & 0 & -(2M+4\eta(a+\omega)) &
0\\
\gamma\sqrt{3}(\omega-a) & -(2M+4\eta(a-\omega)) & 0 &
-{2\eta\overline{\sigma}\sqrt{3}}\\
  -{\sigma\gamma } &
-{2\eta\sigma\sqrt{3}} & 0 & {-{2m}+6\lambda(\omega-a)}
\end{array}}\right) \nonumber \eea

The above matrix is to be diagonalized after imposing the fine
tuning condition $Det {\cal H} =0$ to keep one pair of doublets
light.

\vspace{.5 true cm}

 {\bf{c)}} $[\bar 3,1,{2\over 3}]
(\bt_1,\bt_2,\bt_3,\bt_4,\bt_5) \oplus [3,1,-{2\over 3}]
(t_1,t_2,t_3,t_4,t_5)$\hfil\break $.\qquad\qquad\equiv (H^{\bm
4},\Sigb_{(a)}^{\bm 4}, \s^{\bm 4
}_{(a)},\s^{\bm4}_{R0},\phi_{4(R+)}^{~\bm}) \oplus (H_{\bm
4},\Sigb_{(a)\bm 4}, \s_{\bm 4 (a)},\Sigb_{\bm
4(R0)},\phi_{\bm(R-)}^{~4}) $

\bea {\cal{T}}= \left({\begin{array}{ccccc} M_{H} &
\overline{\gamma}(a+p) & {\gamma}(p-a) & {2\sqrt{2}i
\omega{\bar\gamma}} & i\bar{\sigma}\bar{\gamma}\\ \bar\gamma(p-a)
& 0 & 2M & 0 & 0\\ \gamma(p+a) & 2M & 0 & 4\sqrt{2}i \omega\eta &
2i{\eta\overline{\sigma}}\\ -2\sqrt{2}i \omega\gamma  &
-4\sqrt{2}i \omega\eta & 0 & 2M+2\eta{p}+2\eta a &
-2\sqrt{2}{\eta\overline{\sigma}}\\ i\sigma\gamma & 2i \eta\sigma
& 0 & 2\sqrt{2}\eta\sigma & -2m - 2\lambda(a+p-4 \omega)
\end{array}}\right)
\nonumber\eea

\vspace{.5 cm}
\newpage

 {\bf{iii)  Mixed gauge chiral.}}

{\bf{a)}}$ [1,1,0] (G_1,G_2,G_3,G_4,G_5,G_6) \equiv
(\phi,\phi^{(15)},\phi^{(15)}_{(R0)},{{\s^{44}_{(R-)}}\over \sq},
{{\Sigb_{44((R+)}}\over \sq}, {{{\sq \lambda^{(R0)} -
{\sqrt{3}}\lambda^{(15)}}\over {\sqrt{5}}}})$

\bea {\cal{G}}= 2\left({\begin{array}{cccccc} m&0 &
\sqs\la\om & {{i\e\ssb}\over \sq}&{-i\e\sss\over \sq}&0\\
0& m + 2 \la a & 2\sq\la\om& i\e\ssb\sqtt &-i\e\sss\sqtt&0\\
\sqs\la\om&2\sq\la\om&m+\la(p+2a)& -i\e\sqt\ssb & i\sqt\e\sss&0\\
{{i\e\ssb}\over\sq}& i\e\ssb\sqtt&-i\e\sqt\ssb&0&
M+\e(p+3a -6\om)&{{\sqf g\sss^*}\over  2 }\\
{{-i\e\sss}\over\sq}& -i\e\sss\sqtt&i\e\sqt\sss&
M+\e(p+3a -6\om)&0&{{\sqf g \ssb^*}\over 2}\\
0&0&0&{{\sqf g\sss^*}\over 2}&{{\sqf g\ssb^*}\over 2}&0
\end{array}}\right)
\nonumber\eea

{\bf{b)}}  $[\bar 3,2,-{1\over 3}](\bar E_2,\bar E_3,\bar E_4,\bar
E_5) \oplus [3,2,{1\over 3}](E_2,E_3,E_4,E_5)$\hfil\break
$.\qquad\qquad \equiv (\Sigb_{4\alpha\dot 1}^{\bm}, \phi^{\bm
4}_{(s)\alpha\dot 2}, \phi^{(a) \bm 4}_{\alpha\dot 2},\lambda^{\bm
4}_{\alpha\dot 2}) \oplus  (\s_{\bm\alpha\dot 2}^4,\phi_{\bm
4\alpha\dot 1}^{(s)}, \phi_{\bm 4\alpha\dot
1}^{(a)},\lambda_{\bm\alpha\dot 1}) $

\bea {\cal{E}}= \left({\begin{array}{cccc} -2(M+\e(a-3\om))&
-2\sq i\e\sss&2i\e\sss&ig\sq\ssb^*\\
2i\sq\e\ssb&-2(m+\la(a-\om))&-2\sq\la\om&2g(a^*-\om^*)\\
-2i\e\ssb&-2\sq\la\om&-2(m-\la\om)&\sq g(\om^*-p^*)\\
-ig\sq\sss^*&2g(a^*-\om^*)&g\sq(\om^*-p^*)&0
 \end{array}}\right)
\eea

{\bf{c)}}$ [1,1,-2](\bar F_1,\bar F_2, \bar F_3)
 \oplus [1,1,2](F_1,F_2,F_3) $ \hfil\break
$.\qquad\qquad \equiv
(\Sigb_{44(R0)},\phi^{(15)}_{(R-)},\lambda_{(R-)})
  \oplus (\s^{44}_{(R0)},\phi^{(15)}_{(R+)},\lambda_{(R+)}) $.

\bea {\cal{F}}= \left({\begin{array}{ccc} 2(M+\e(p+3 a))&
-2i\sqt\e\sss&-g\sq\ssb^*\\
2i\sqt\e\ssb&2(m+\la(p+2a))& {\sqrt{24}}ig\om^*)\\
 -g\sq\sss^*&-{\sqrt{24}}ig\om^*&0
 \end{array}}\right)
\eea

{\bf{d)}} $[\bar 3,1,-{4\over 3}](\bar J_1,\bar J_2,\bar J_3,\bar
J_4) \oplus [3,1,{4\over 3}](J_1,J_2,J_3,J_4)$ \hfil\break
$.\qquad\qquad\equiv (\s^{\bm4}_{(R-)},\phi_4^{\bm},
\phi_4^{~\bm(R0)},\lambda_4^{~\bm}) \oplus
(\Sigb_{\bm4(R+)},\phi_{~\bm}^4,
\phi_{\bm(R0)}^{~4},\lambda_{\bm}^4)$

\bea {\cal{J}}= \left({\begin{array}{cccc} 2(M+\e(a+p-2\om))&
-2\e\ssb&2\sq \e\ssb&-ig\sq\sss^*\\
2\e\sss&-2(m+\la a)&-2\sq\la\om&-2ig\sq a^*\\
-2\sq\e\sss&-2\sq\la\om&-2(m+\la(a+p))&-4i g\om^*\\
-ig\sq\ssb^*&2\sq ig a^*&4i g\om^*&0
 \end{array}}\right)
\eea

{\bf{e)}}$ [3,2,{5\over 3}](\bar X_1,\bar X_2,\bar X_3) \oplus
[3,2,-{5\over 3}](X_1,X_2,X_3)\hfil\break.\qquad\qquad \equiv
(\phi^{(s)\bm4}_{\alpha\dot 1}, \phi^{(a)\bm4}_{\alpha\dot 1}
,\lambda^{\bm4}_{\alpha\dot 1}) \oplus(\phi_{\bm4\alpha\dot
2}^{(s)}, \phi_{\bm4\alpha\dot 2}^{(a)},\lambda_{\bm4\alpha\dot
2})
  $

\bea {\cal{X}}= \left({\begin{array}{ccc} 2(m+\la(a+\om))&
-2\sq \la \om &-2g(a^*+\om^*)\\
-2\sq \la \om &2(m+\la \om)& {\sq}g(\om^* +p^*)\\
 -2 g(a^* +\om^*) &\sq g(\om^* + p^*)&0
 \end{array}}\right)
\eea

\begin{figure}[h!]
\begin{center}
\epsfxsize15cm\epsffile{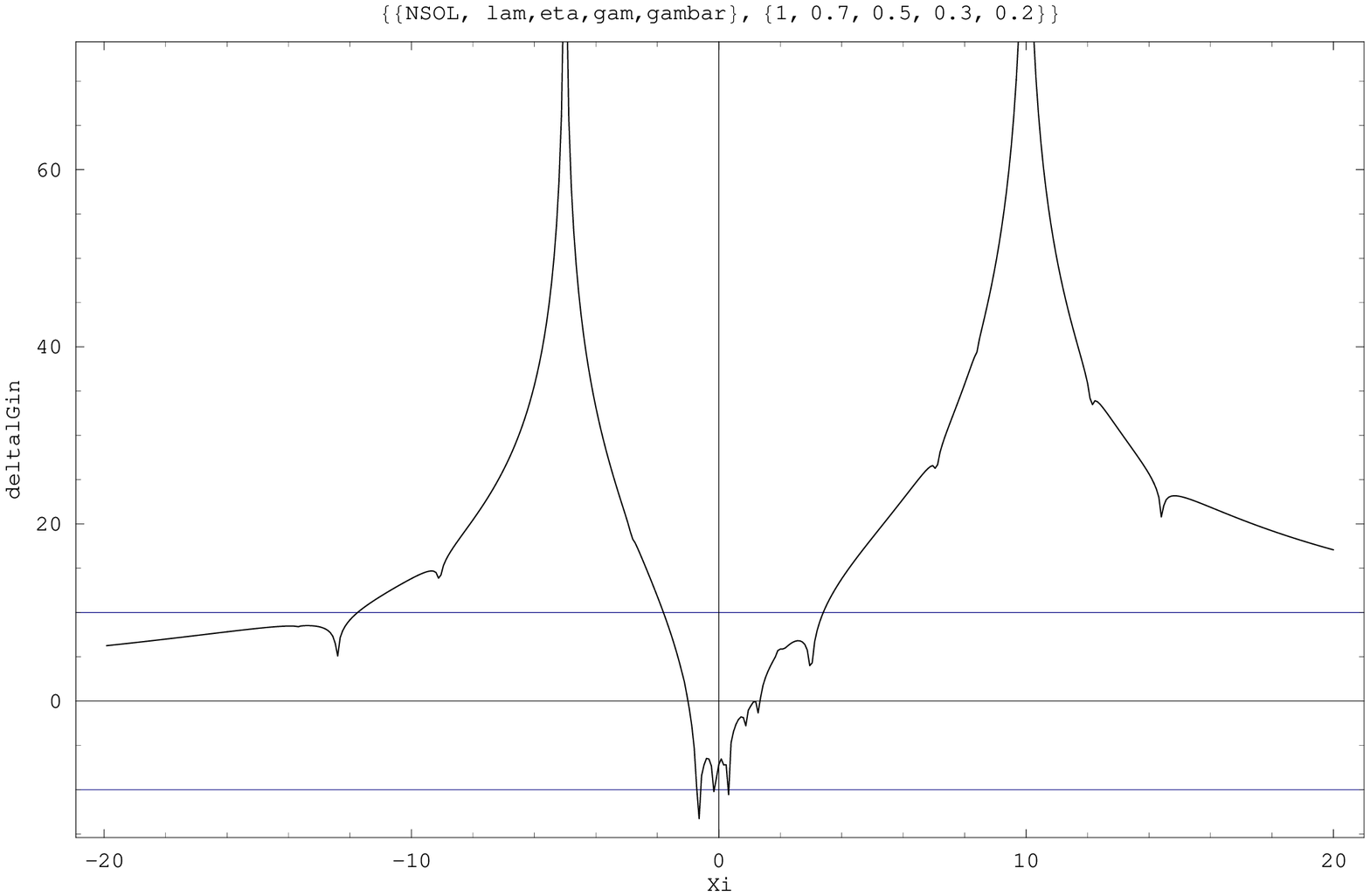}
\epsfxsize15cm\epsffile{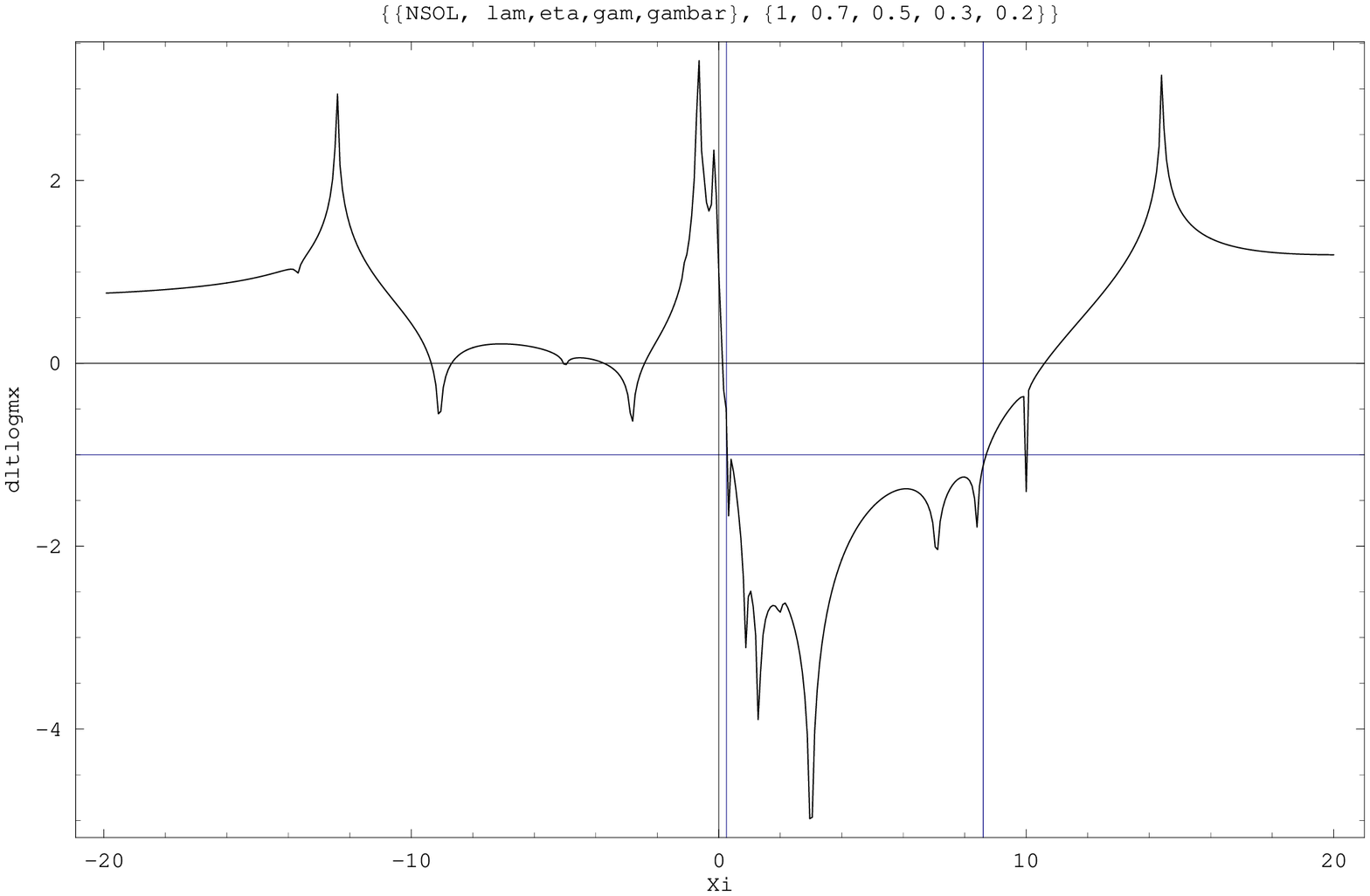}
 \caption{  Plot of the threshold
  corrections $\Delta^{(th)} (\alpha_G^{-1}(M_X))$
  and $\Delta^{(th)}(Log_{10}{M_X})$  vs $\xi$ for real $\xi$ : real
solution for x, $\lambda=.7$. Note the exclusion  of $.25<\xi<8.6$ from the requirement
that  $\Delta^{(th)}(Log_{10}{M_X}) > -1$.}
\end{center}
\end{figure}

\begin{figure}[h!]
\begin{center}
\epsfxsize15cm\epsffile{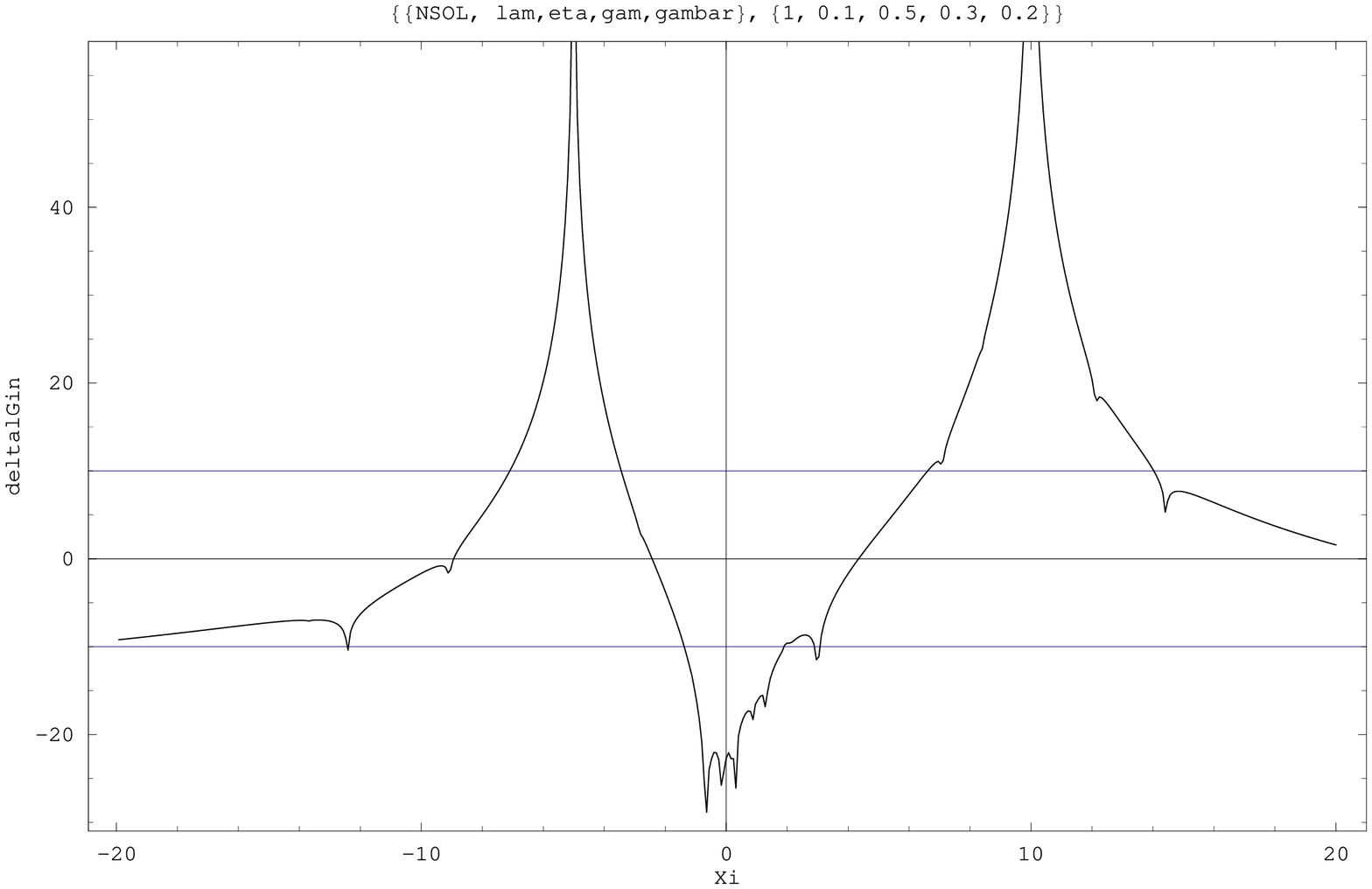}
\epsfxsize15cm\epsffile{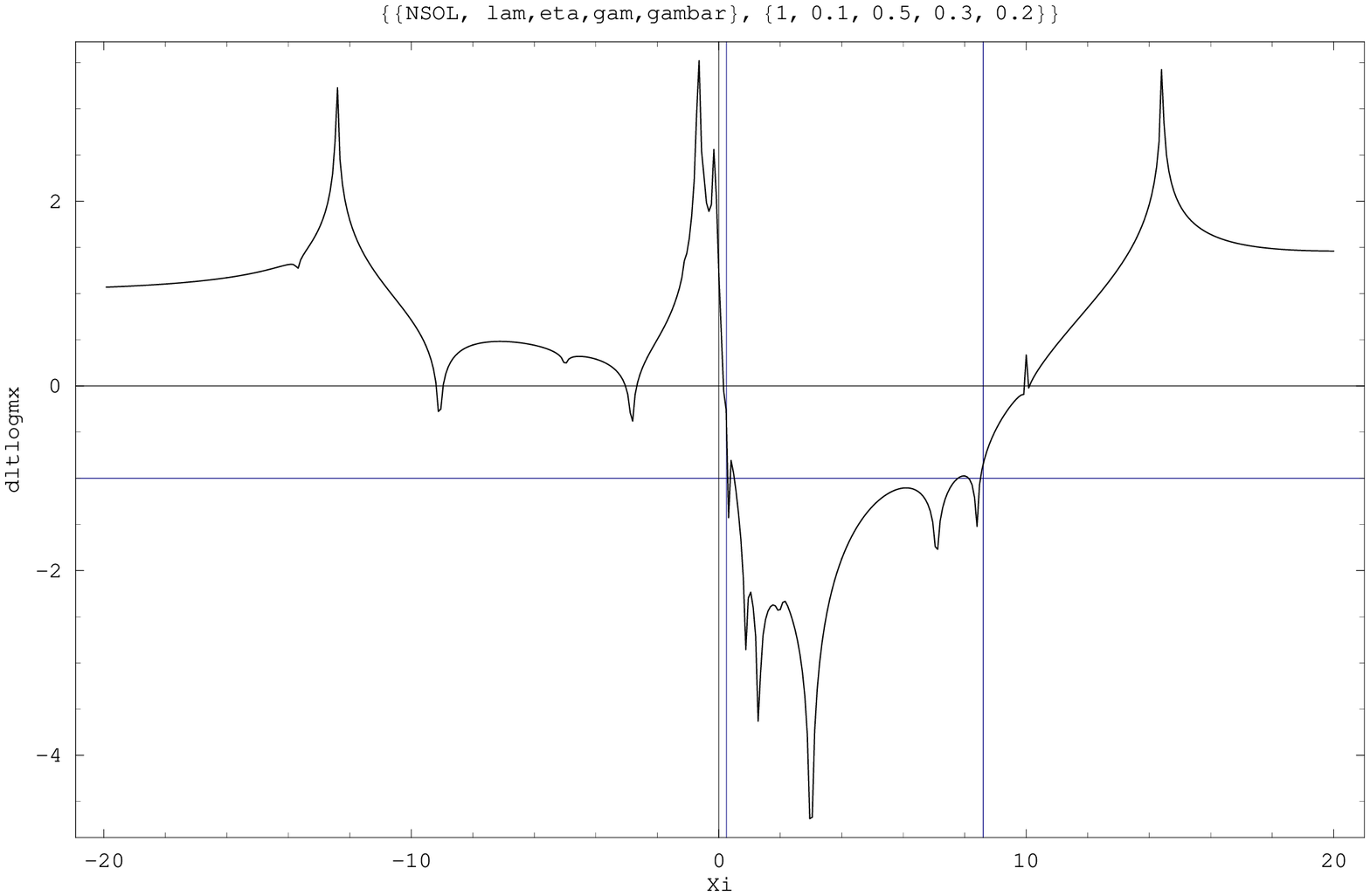}
 \caption{  Plot of the
  corrections $\Delta^{(th)} (\alpha_G^{-1}(M_X))$  and  $\Delta^{(th)}(Log_{10}{M_X})$
vs $\xi$ for real $\xi$ : real solution for x,  $\lambda =.1$. Note how the region
around $\xi=0$ is excluded by   $\Delta^{(th)} (\alpha_G^{-1}(M_X))> -10$
 while  $\Delta^{(th)}(Log_{10}{M_X}) >-1$ continues to rule out the
$8.6 >\xi >.25 $ region.}
\end{center}
\end{figure}

\begin{figure}[h!]
\begin{center}
\epsfxsize15cm\epsffile{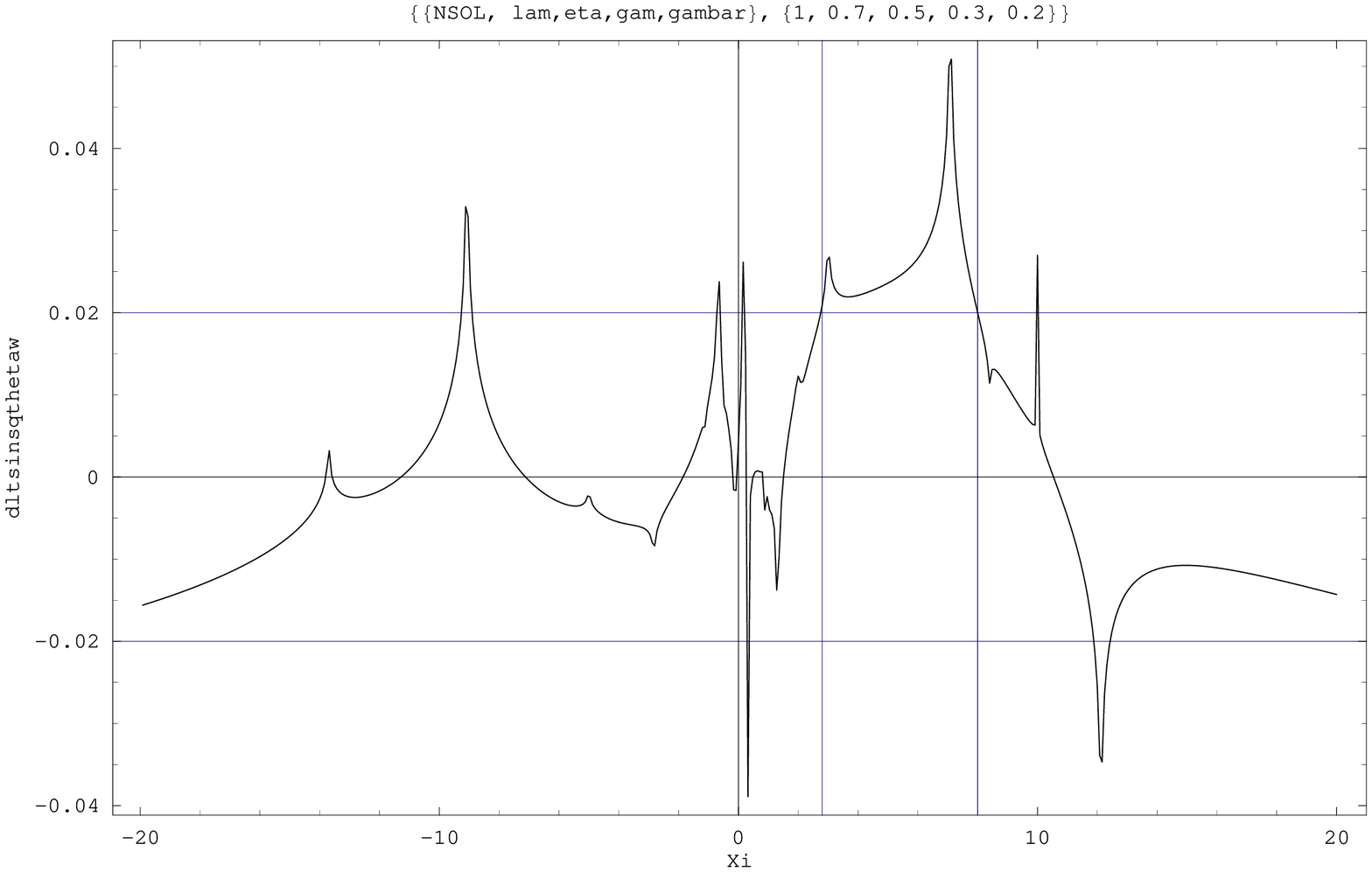}
\epsfxsize15cm\epsffile{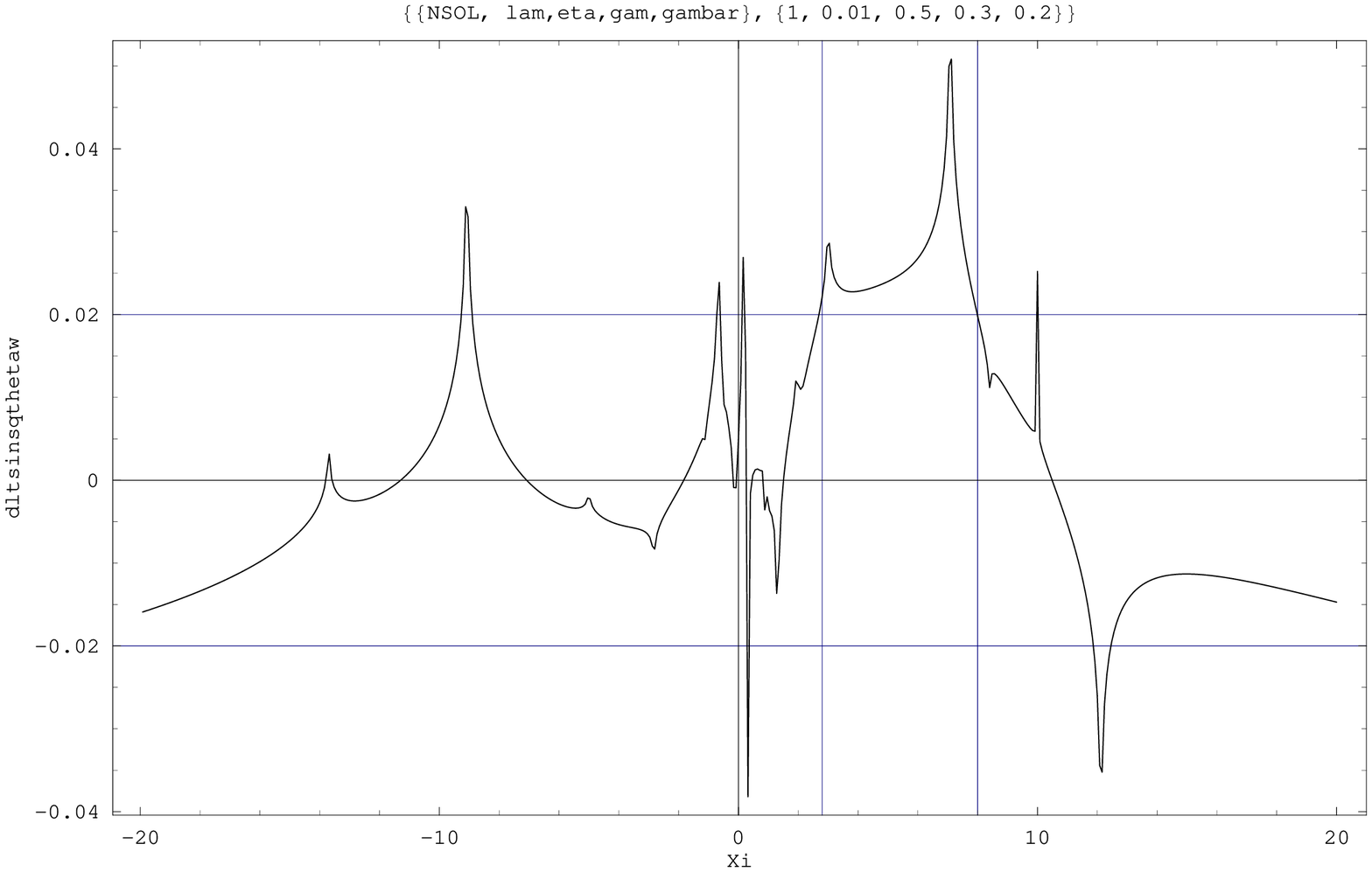}
 \caption{  Plot of
  $\Delta^{(th)} (Sin^2\theta_W (M_S))$
 vs $\xi$ for real $\xi$ : real
solution for x,  $\lambda =.7$ (upper) and $\lambda =.01$ (lower).
  Note how the large change in $Sin^2\theta_W (M_S) $ for $2.8<\xi <8$ also excludes this region.}
\end{center}
\end{figure}

\begin{figure}[h!]
\begin{center}
\epsfxsize15cm\epsffile{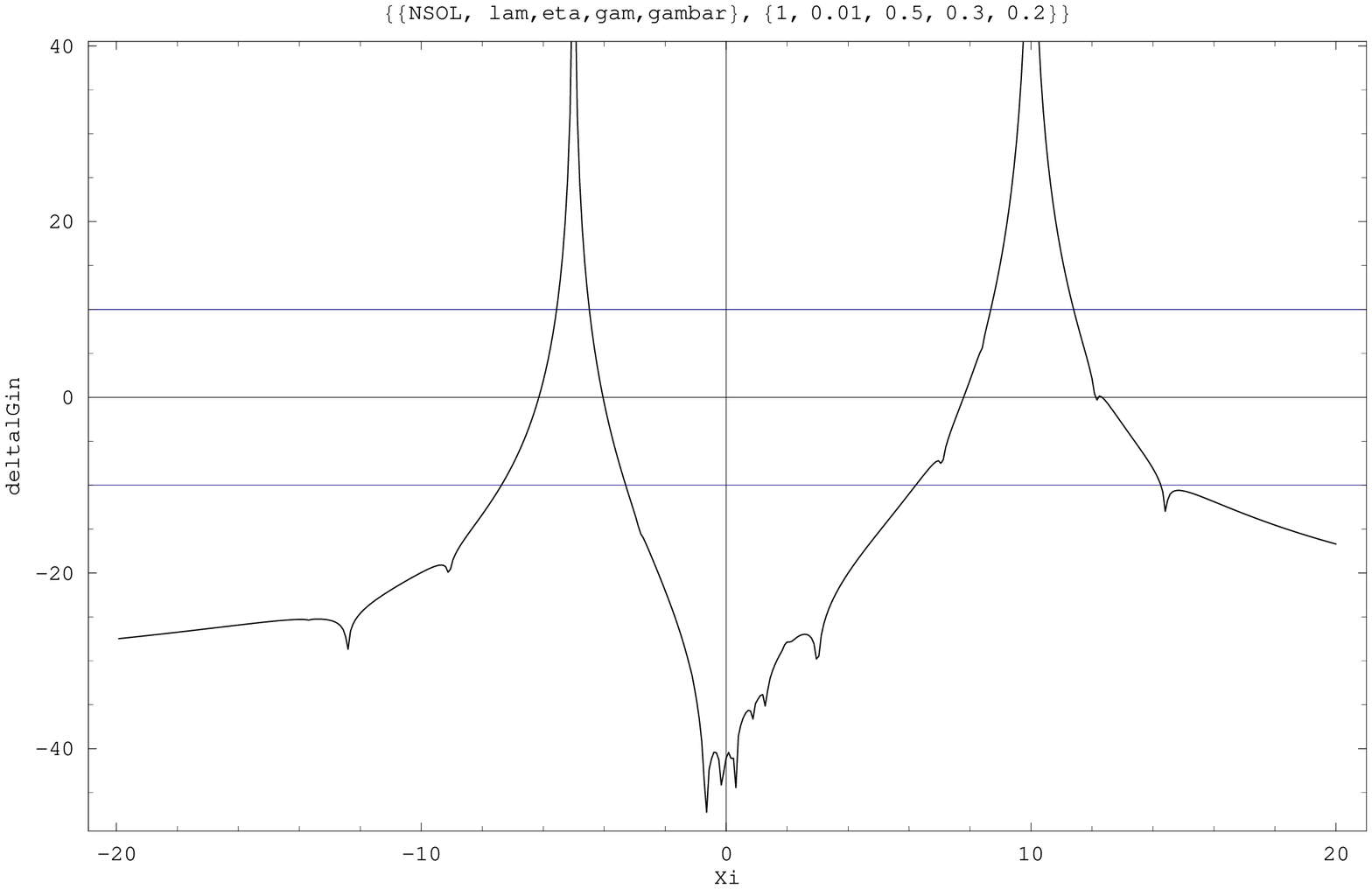}
\epsfxsize15cm\epsffile{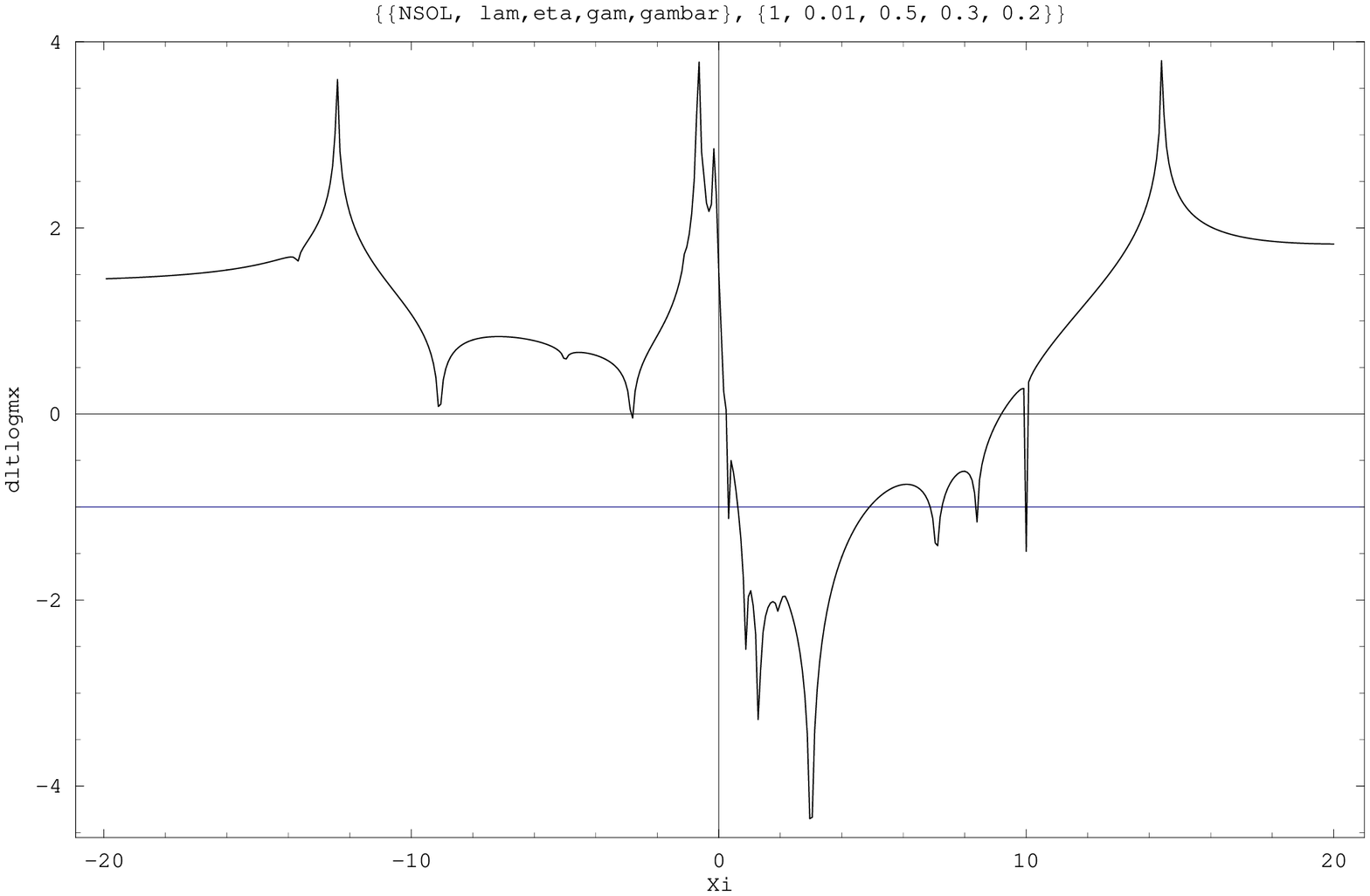}
 \caption{  Plot of the  corrections $\Delta^{(th)} (\alpha_G^{-1}(M_X))$  and
$\Delta^{(th)}(Log_{10}{M_X})$ vs $\xi$ for real $\xi$ : real
solution for x, $\lambda =.01$. Only intermediate values of $\xi$ around the two $SU(5)$ singularities
are now viable since $\alpha_G$ is too large.}
\end{center}
\end{figure}

\begin{figure}[h!]
\begin{center}
\epsfxsize15cm\epsffile{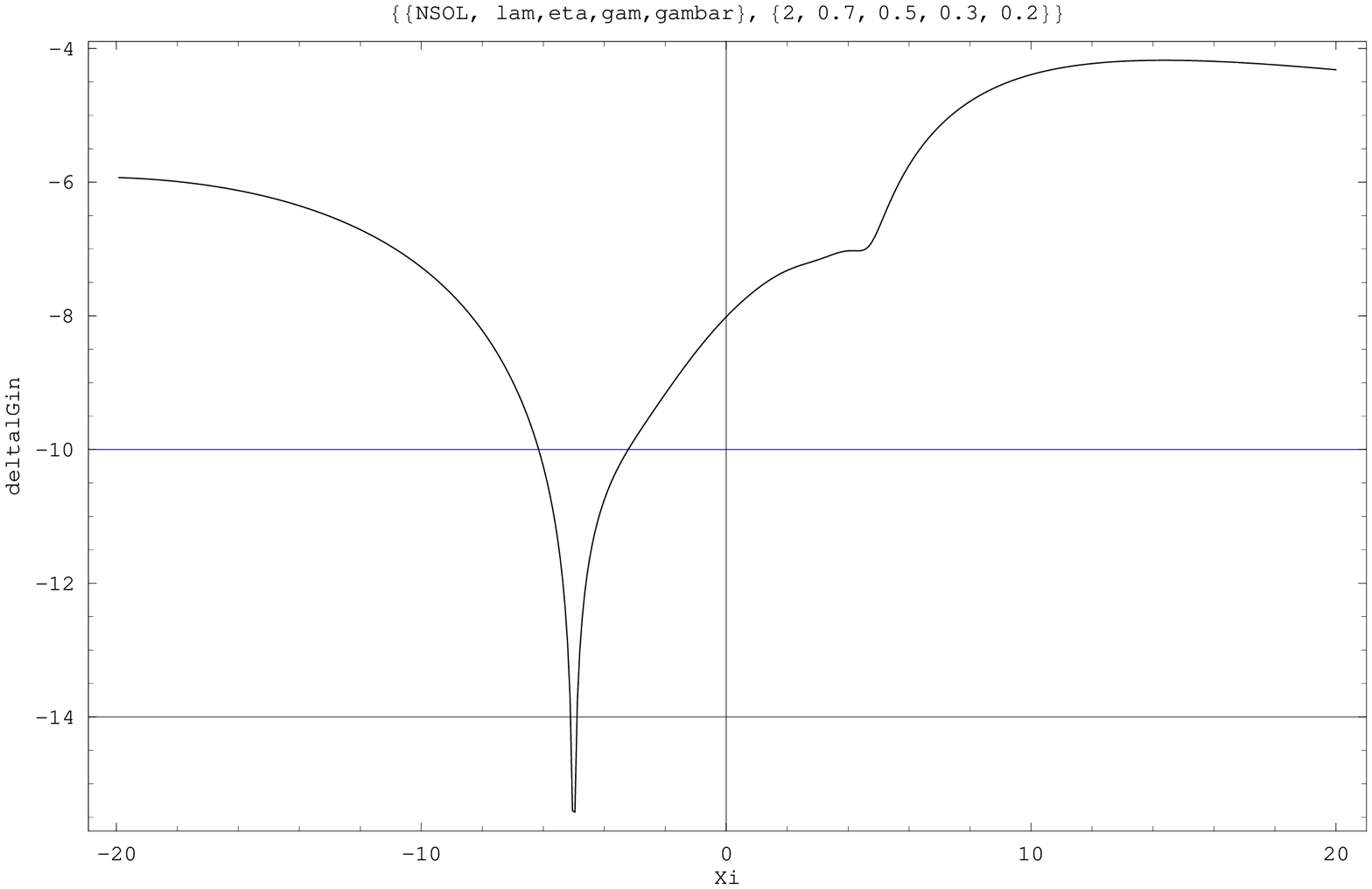}
\epsfxsize15cm\epsffile{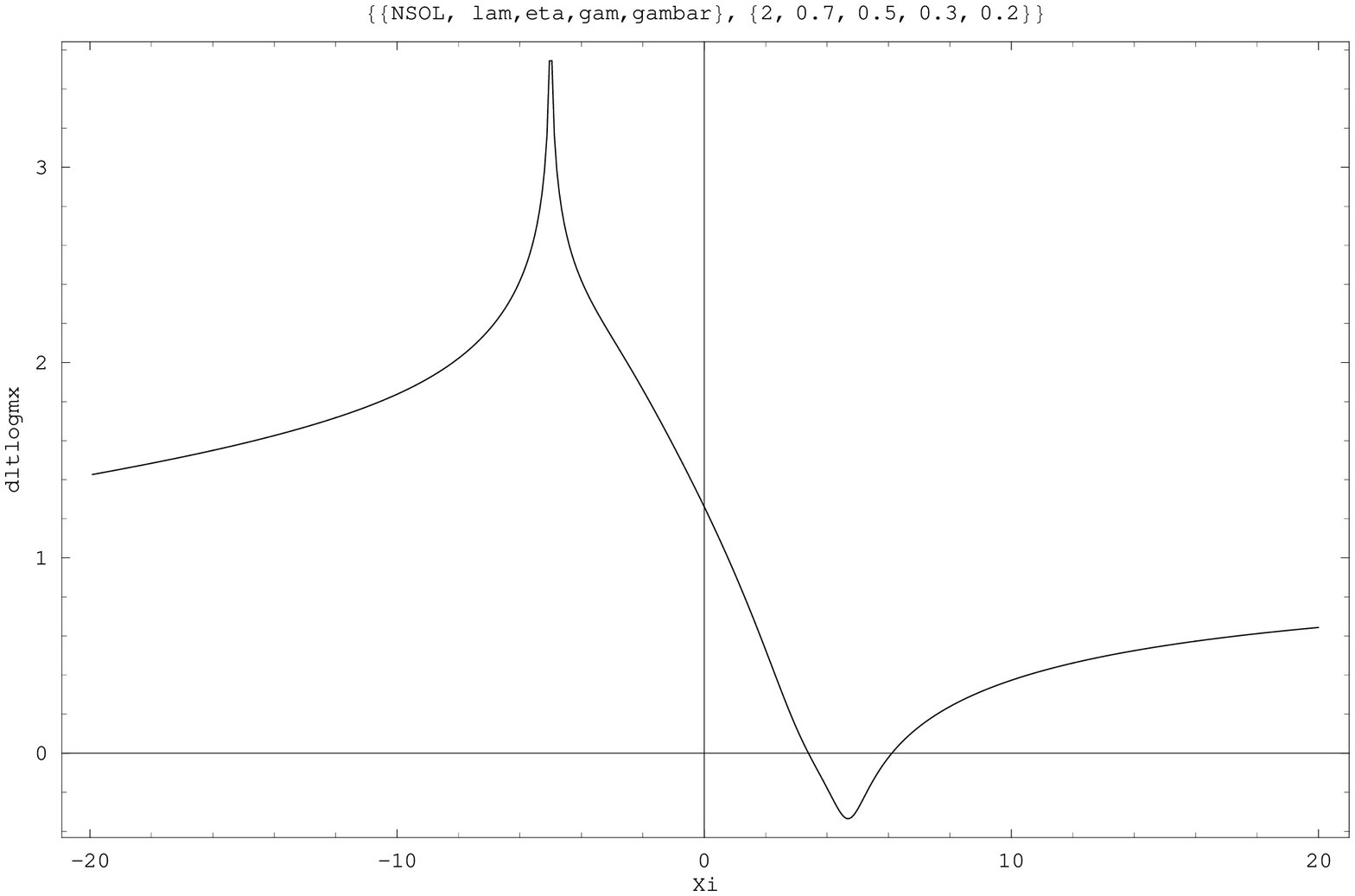}
 \caption{ Plot of the corrections $\Delta^{(th)} (\alpha_G^{-1}(M_X))$  and
$\Delta^{(th)}(Log_{10}{M_X})$      vs $\xi$ for real $\xi$ :  complex
solution for x,    $\lambda =.7$.   Most values of $\xi$ are allowed.}
\end{center}
\end{figure}

\begin{figure}[h!]
\begin{center}
\epsfxsize15cm\epsffile{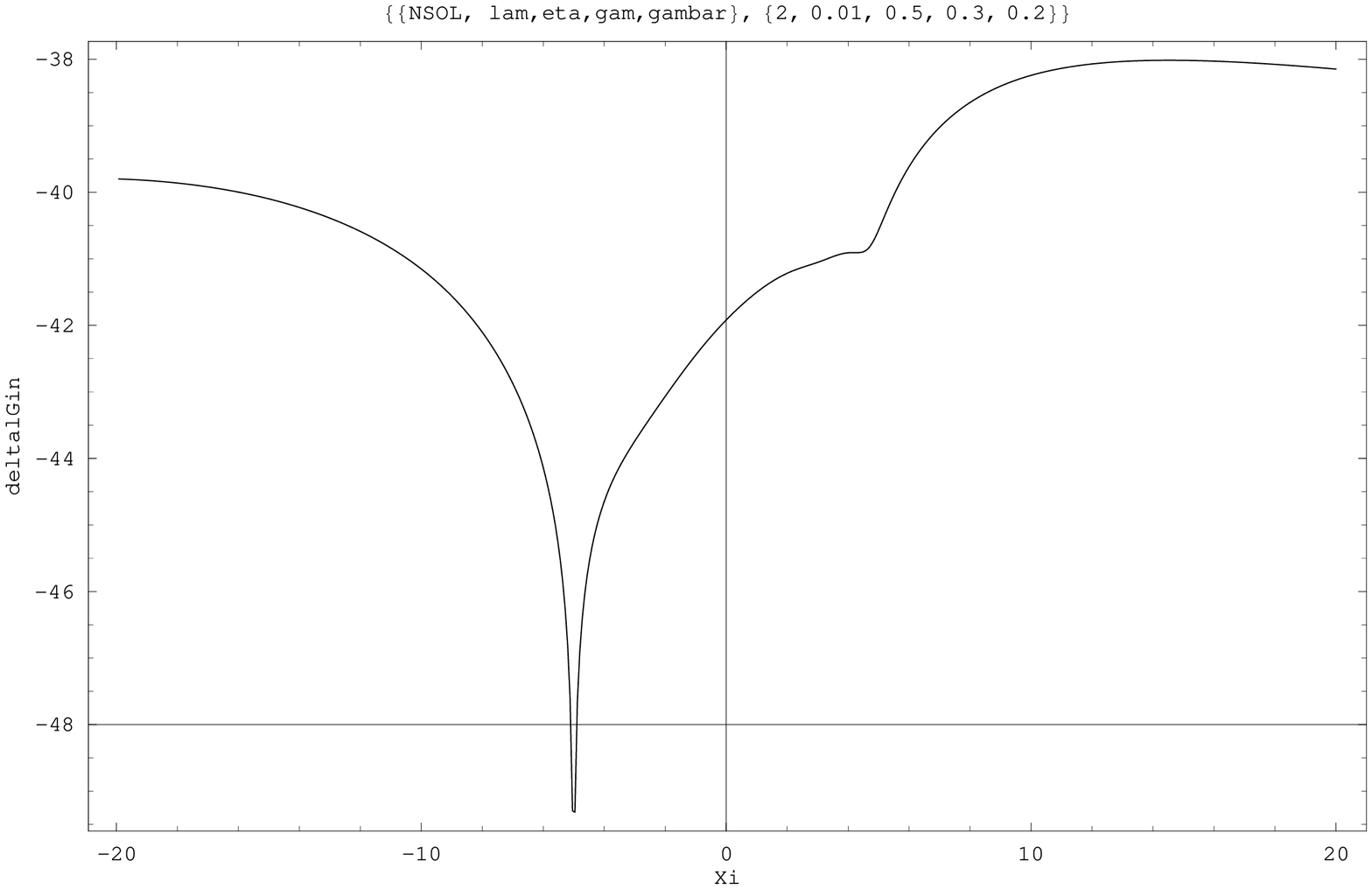}
\epsfxsize15cm\epsffile{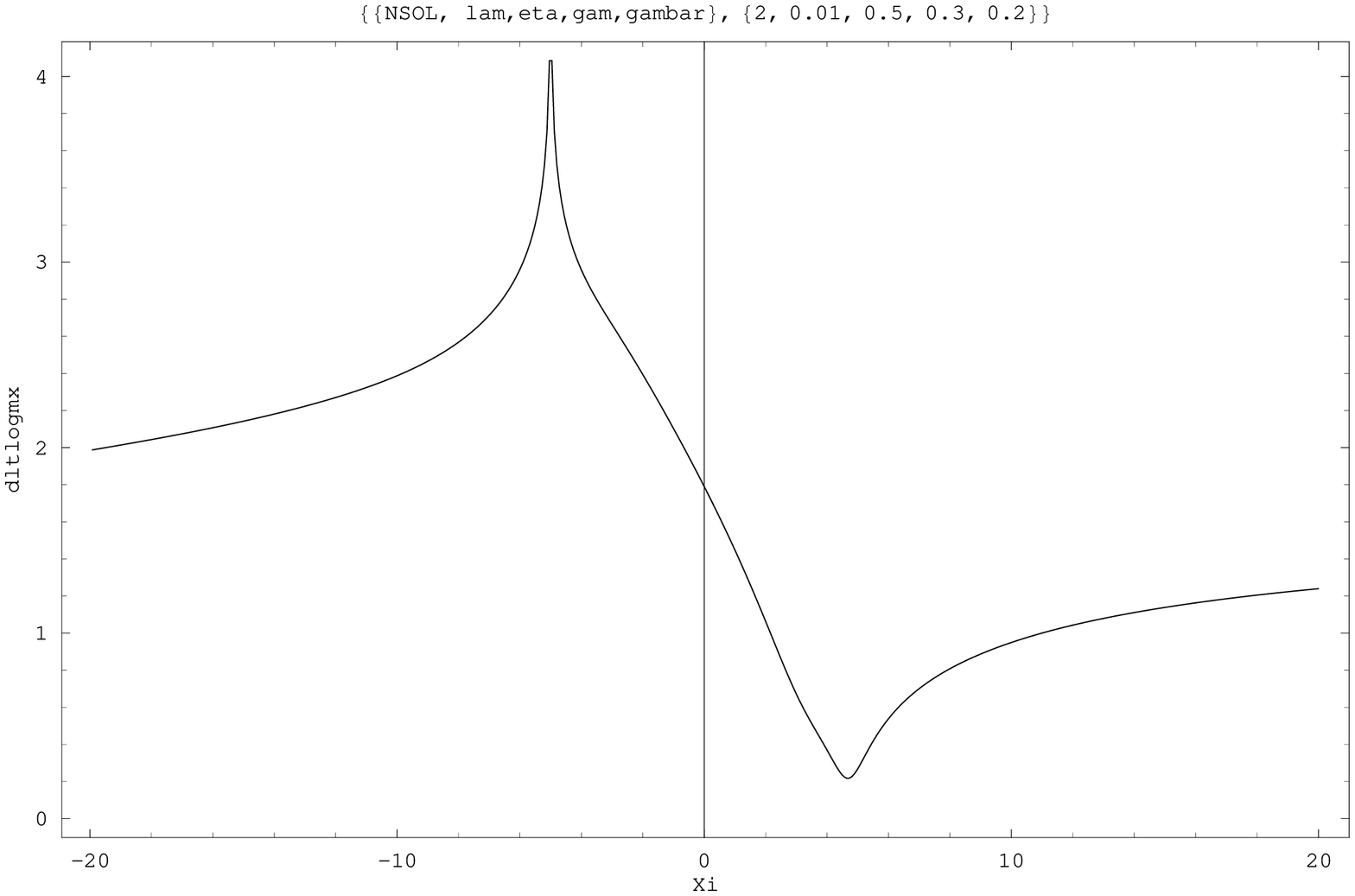}
 \caption{ Plot of the  corrections $\Delta^{(th)} (\alpha_G^{-1}(M_X))$  and
$\Delta^{(th)}(Log_{10}{M_X})$  vs $\xi$ for real $\xi$ :  complex
solution for x,    $\lambda =.01$. Decreasing $\lambda$  further caused
 a catasrophic increase of  $\alpha_G (M_X) $}
\end{center}
\end{figure}

\begin{figure}[h!]
\begin{center}
\epsfxsize15cm\epsffile{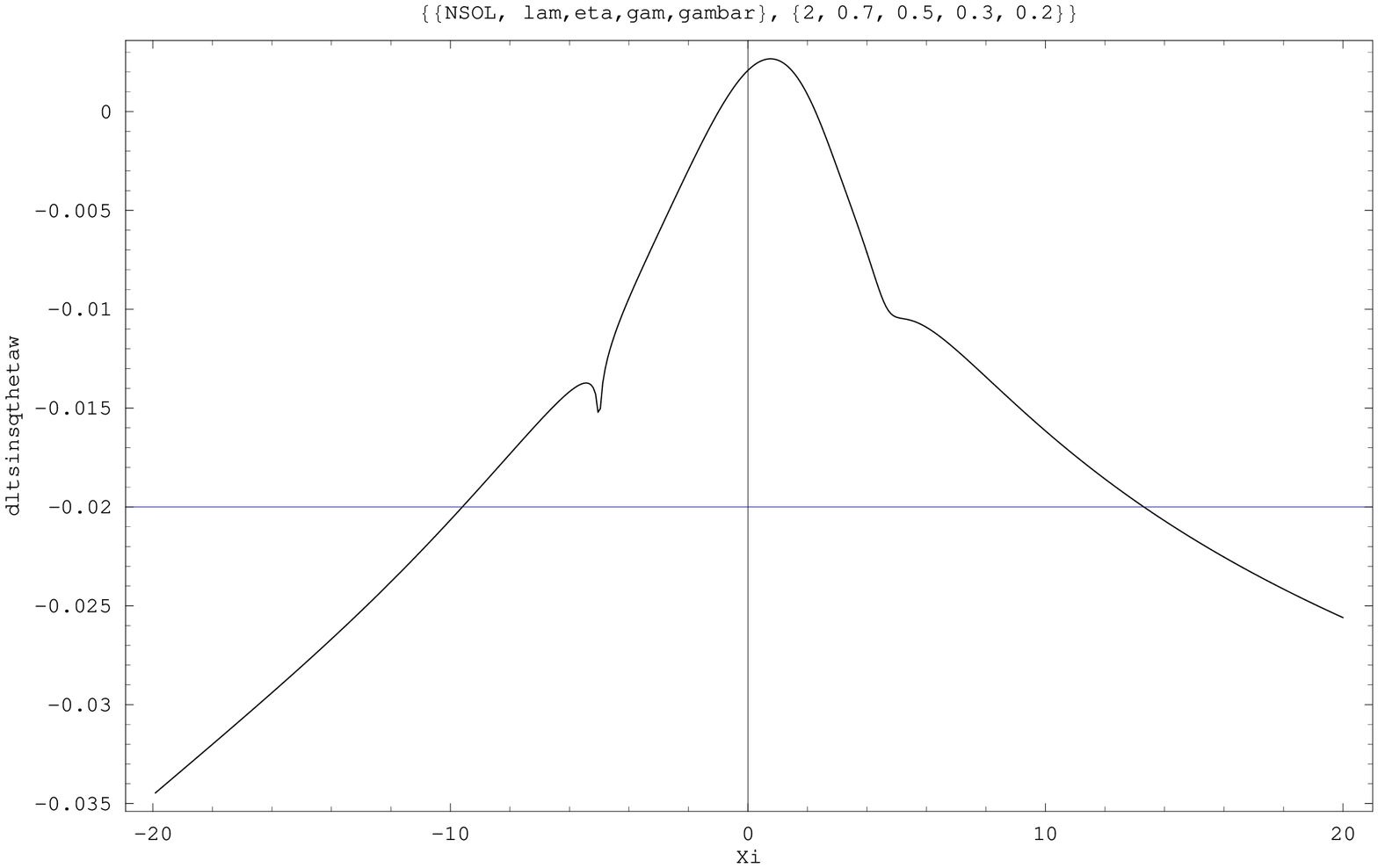}
\epsfxsize15cm\epsffile{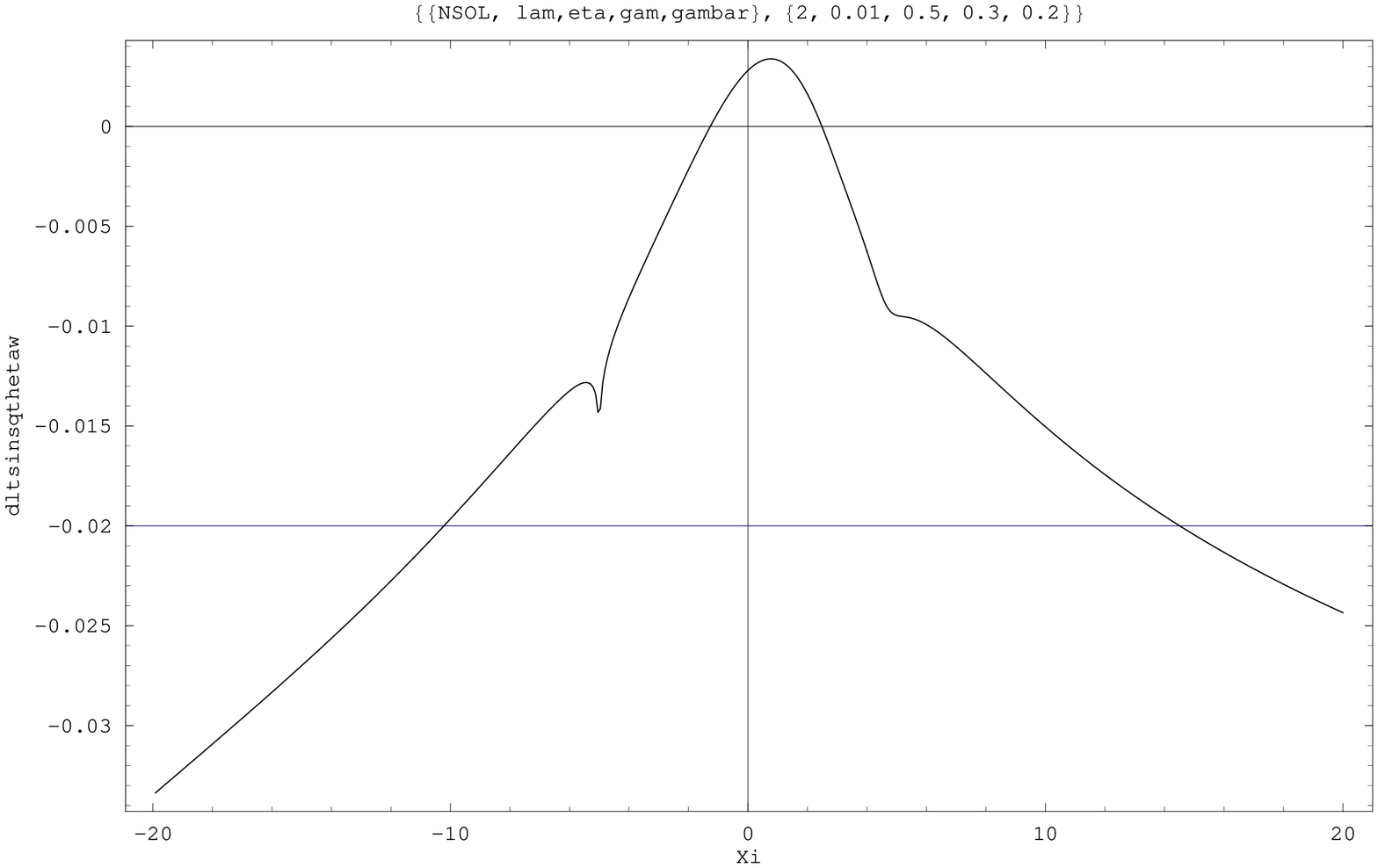}
 \caption{  Plot of the threshold
and two-loop corrections  to  $Sin^2\theta_W (M_S))$
 vs $\xi$ for real $\xi$ :  complex
solution for x,  $\lambda =.7$ and $\lambda =.01$. Values of $|\xi| < 10 $  are allowed. }
\end{center}
\end{figure}

\begin{figure}[h!]
\begin{center}
\epsfxsize15cm\epsffile{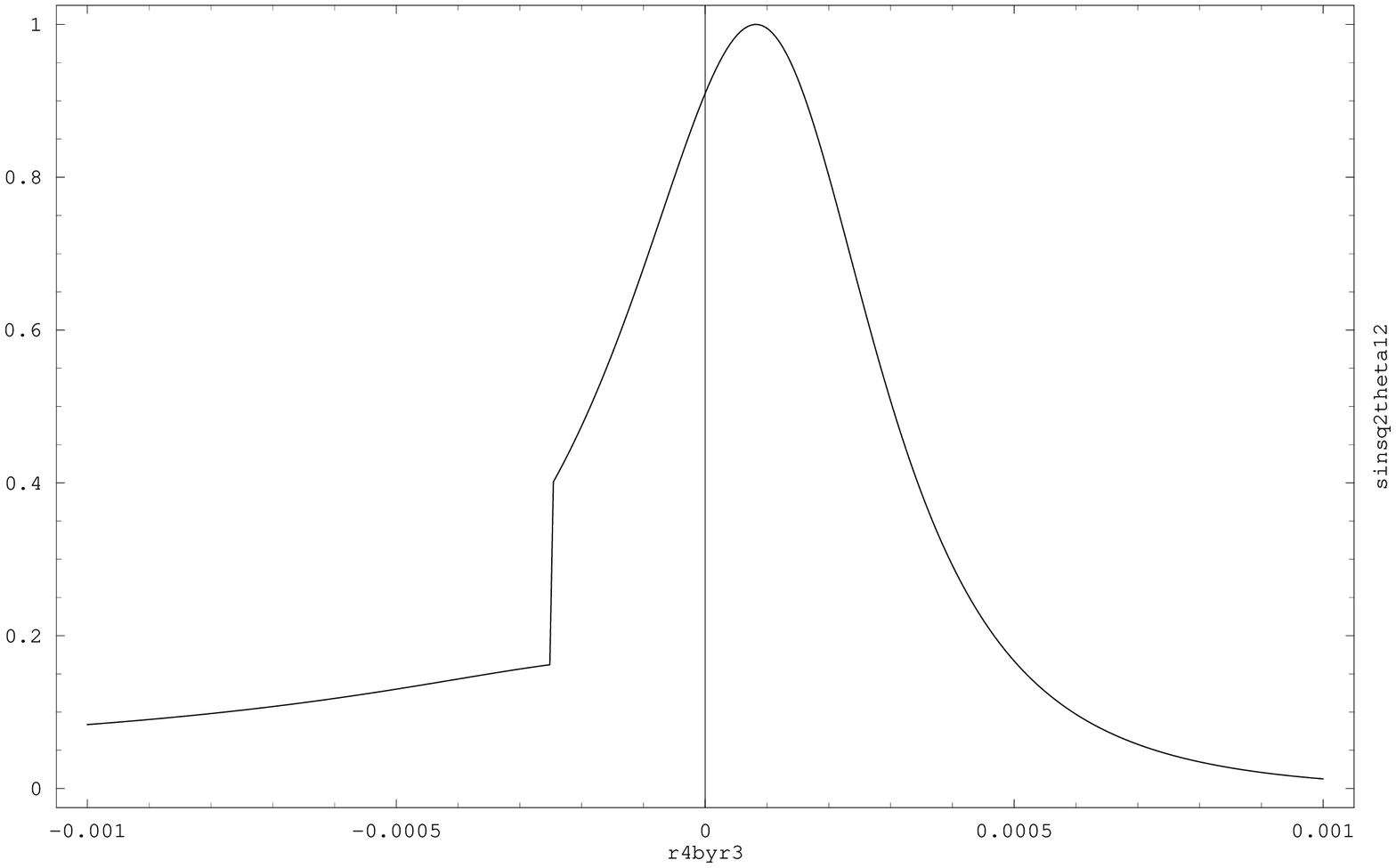}
\epsfxsize15cm\epsffile{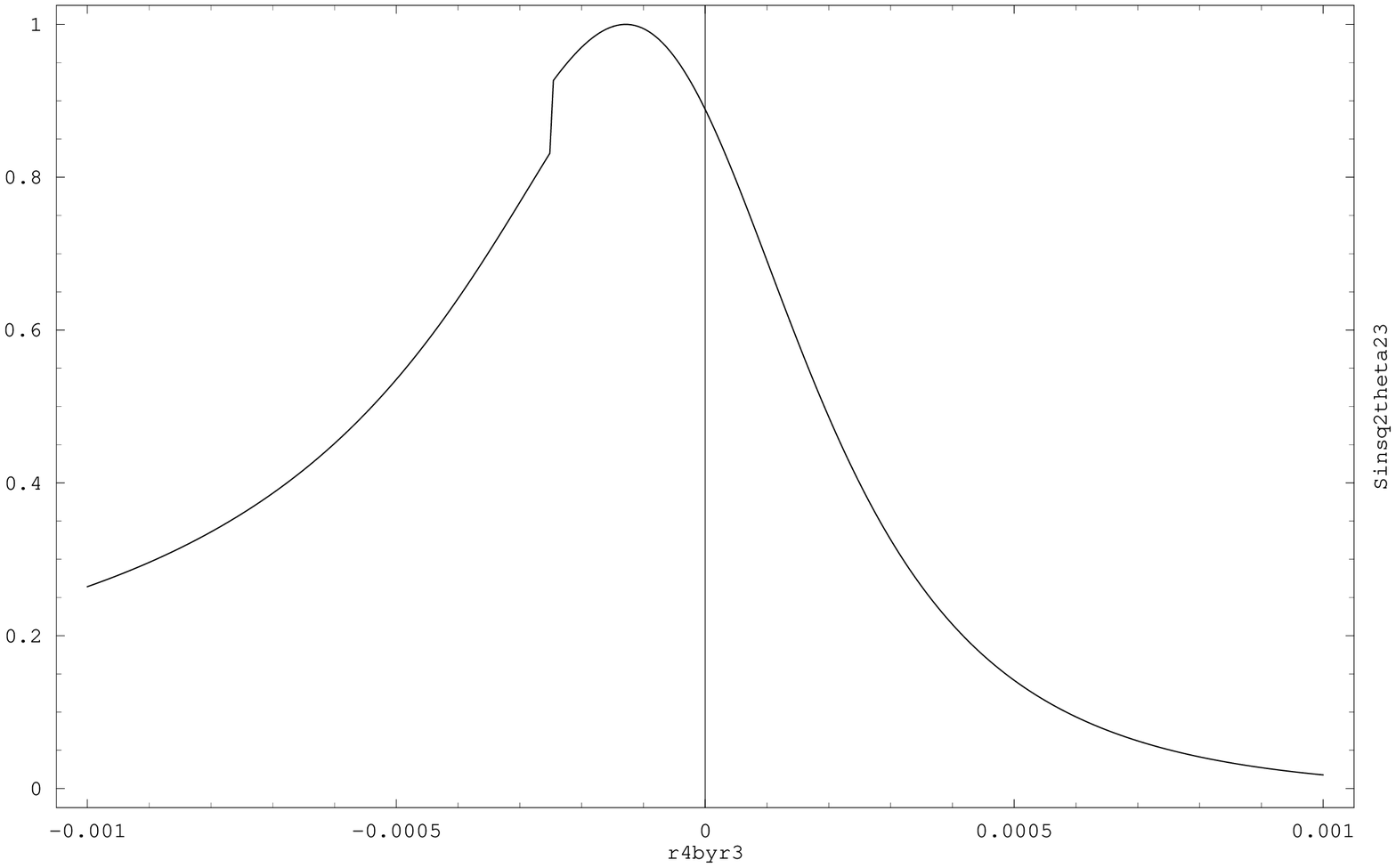}
 \caption{    $ Sin^2 2\theta_{12}, Sin^2 2\theta_{23}$
 vs $r_4/r_3$  for the sample real solution of Bertolini and Malinsky.
 Note the rapid collapse of the large Type II angles as $|r_4/r_3|$ increases.    }
\end{center}
\end{figure}

\begin{figure}[h!]
\begin{center}
\epsfxsize15cm\epsffile{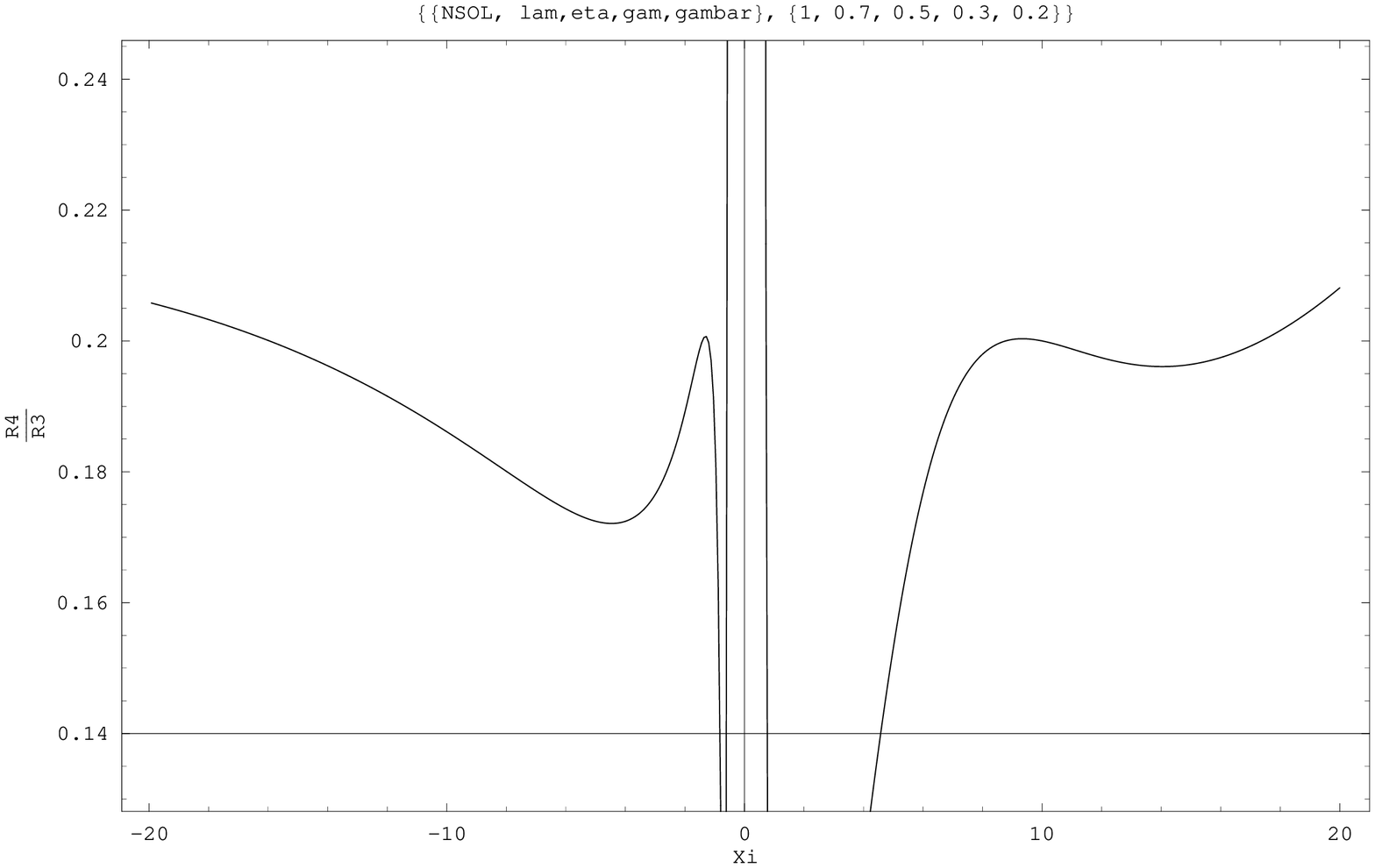}
\epsfxsize15cm\epsffile{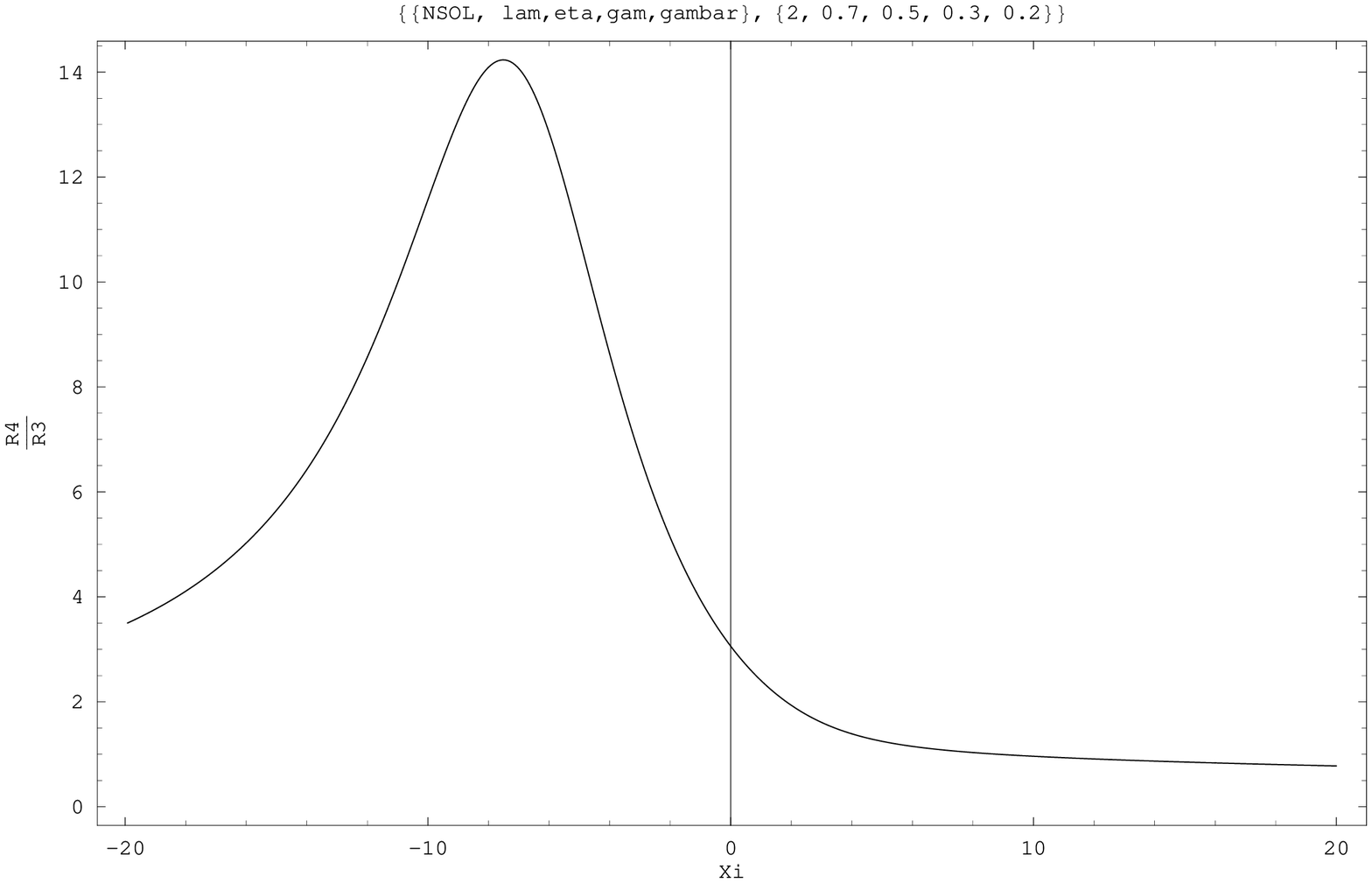}
\caption{   Plots of  MSGUT $R=r_4/r_3$ vs $\xi$ for real $\xi$ :  real
solution(upper)  and complex solution(lower),  $\lambda =.7$. In the real case
the candidate Type II region  $.8 <\xi <4$  with small  R  is  disallowed  by
 $\Delta^{(th)}(Log_{10}{M_X}) >-1 $.  In the complex case  there is no Type II candidate region.}
\end{center}
\end{figure}

\begin{figure}[h!]
\begin{center}
\epsfxsize15cm\epsffile{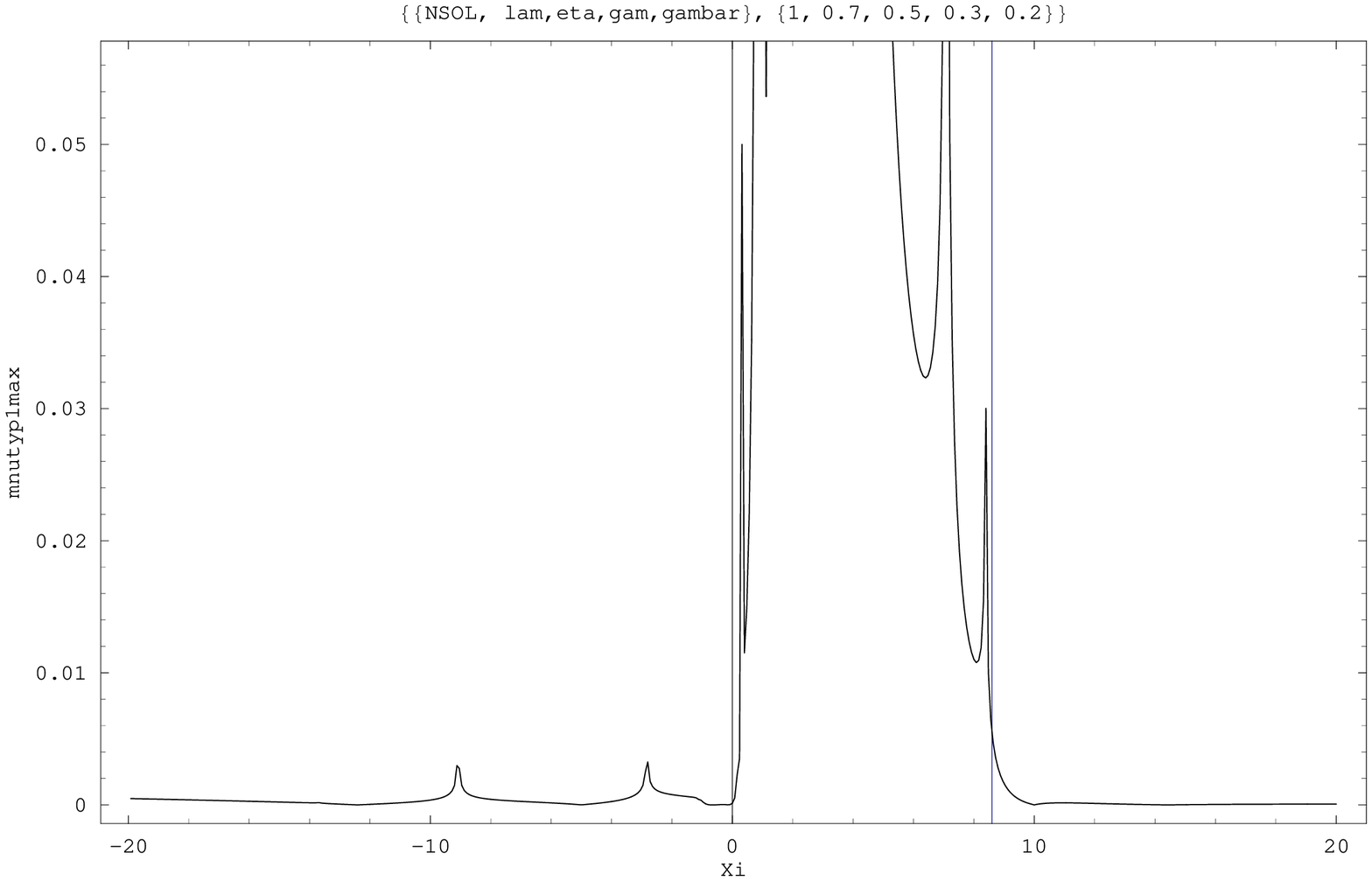}
\epsfxsize15cm\epsffile{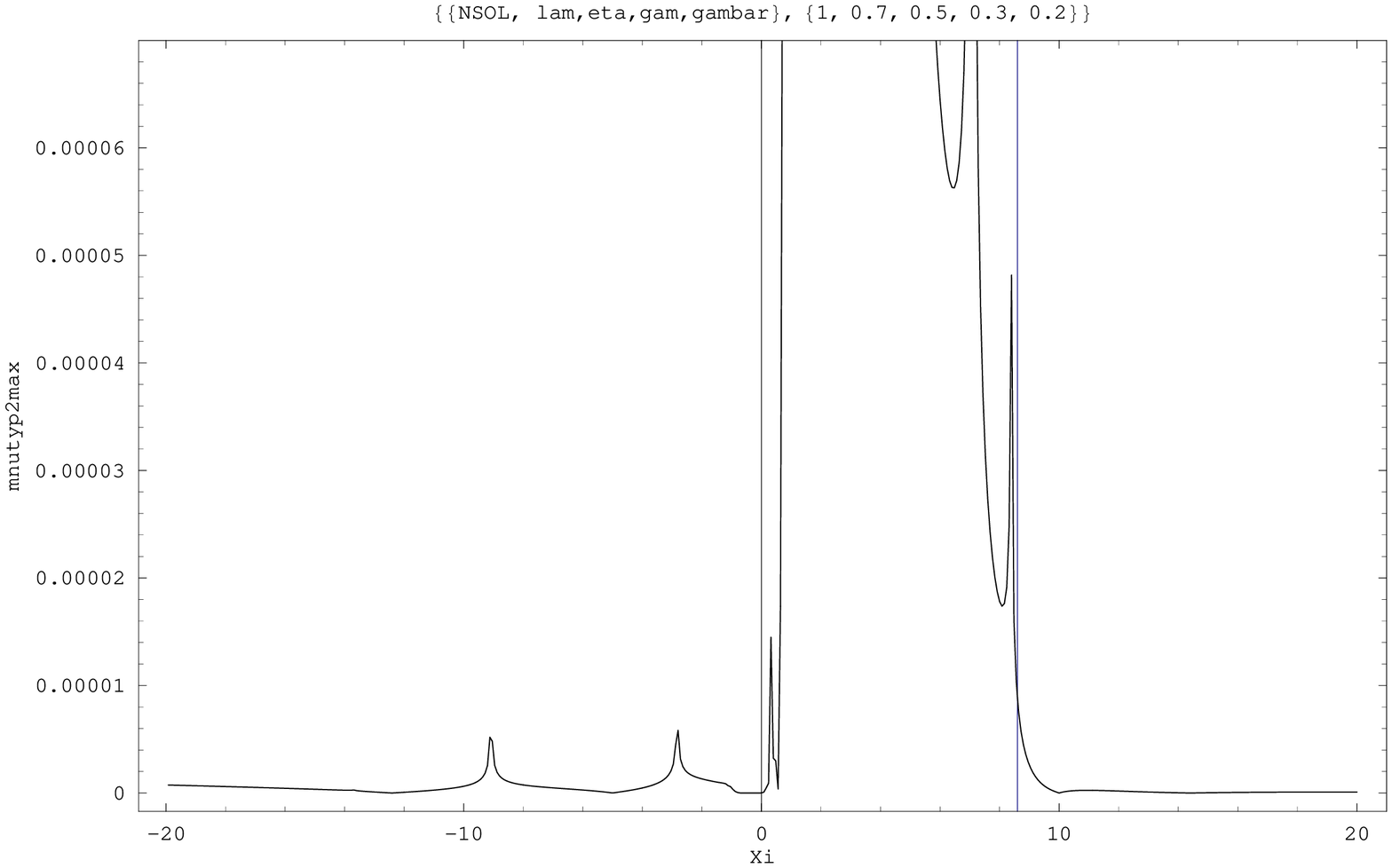}
\caption{   Plots of  the maximum neutrino masses in the MSGUT
(obtained using the sample solution of Bertolini and Malinsky)  vs $\xi$ for real $\xi$ :  real
solution     for x,  $\lambda =.7$.   Type I  (upper) and Type II (lower).
The candidate region $.25<\xi <8.6$  with  possibly large $m_{\nu}^{max}$
  is  disallowed  by $\Delta^{(th)}(Log_{10}{M_X}) >-1 $.}
\end{center}
\end{figure}

  \begin{figure}[h!]
\begin{center}
\epsfxsize15cm\epsffile{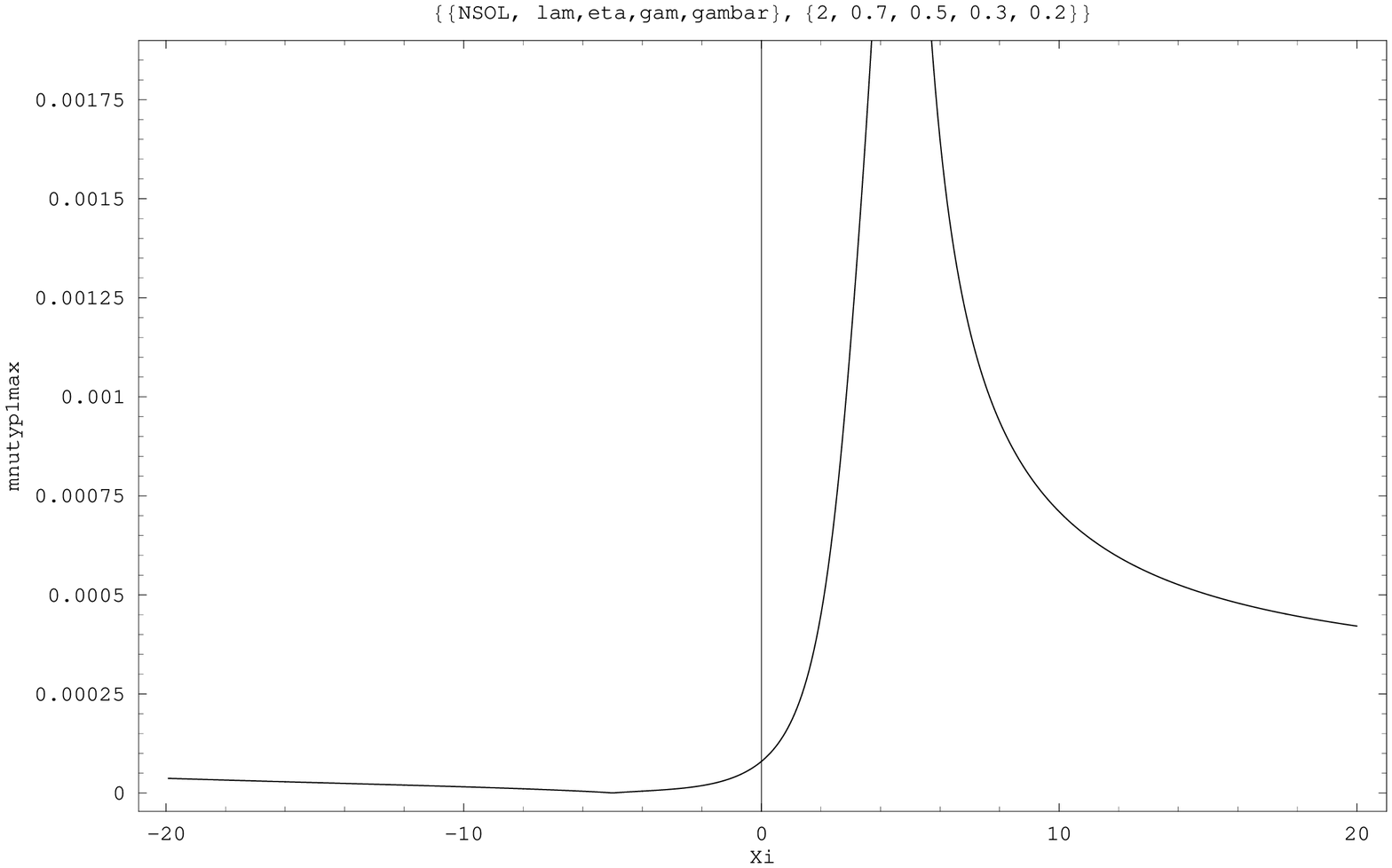}
\epsfxsize15cm\epsffile{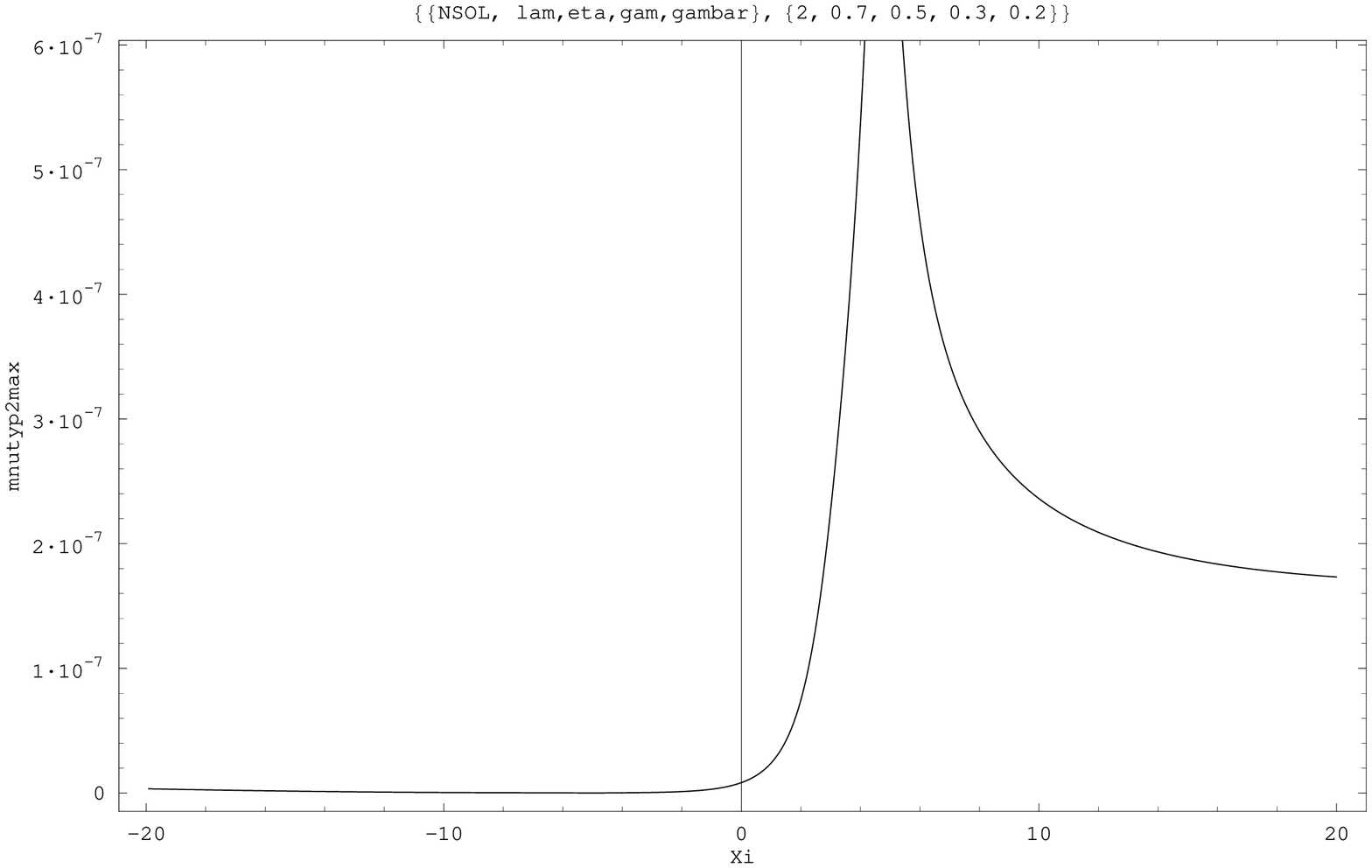}
 \caption{   Plots of  the maximum neutrino masses in the MSGUT
(obtained using the sample solution of Bertolini and Malinsky)   vs $\xi$ for real $\xi$ :
  complex solution   for x,  $\lambda =.7$.  Type I  (upper) and Type II (lower).
Type I solutions are always dominant
and may have the right magnitude only in $4<\xi<6$   region.  }
\end{center}
\end{figure}

\begin{figure}[h!]
\begin{center}
\epsfxsize15cm\epsffile{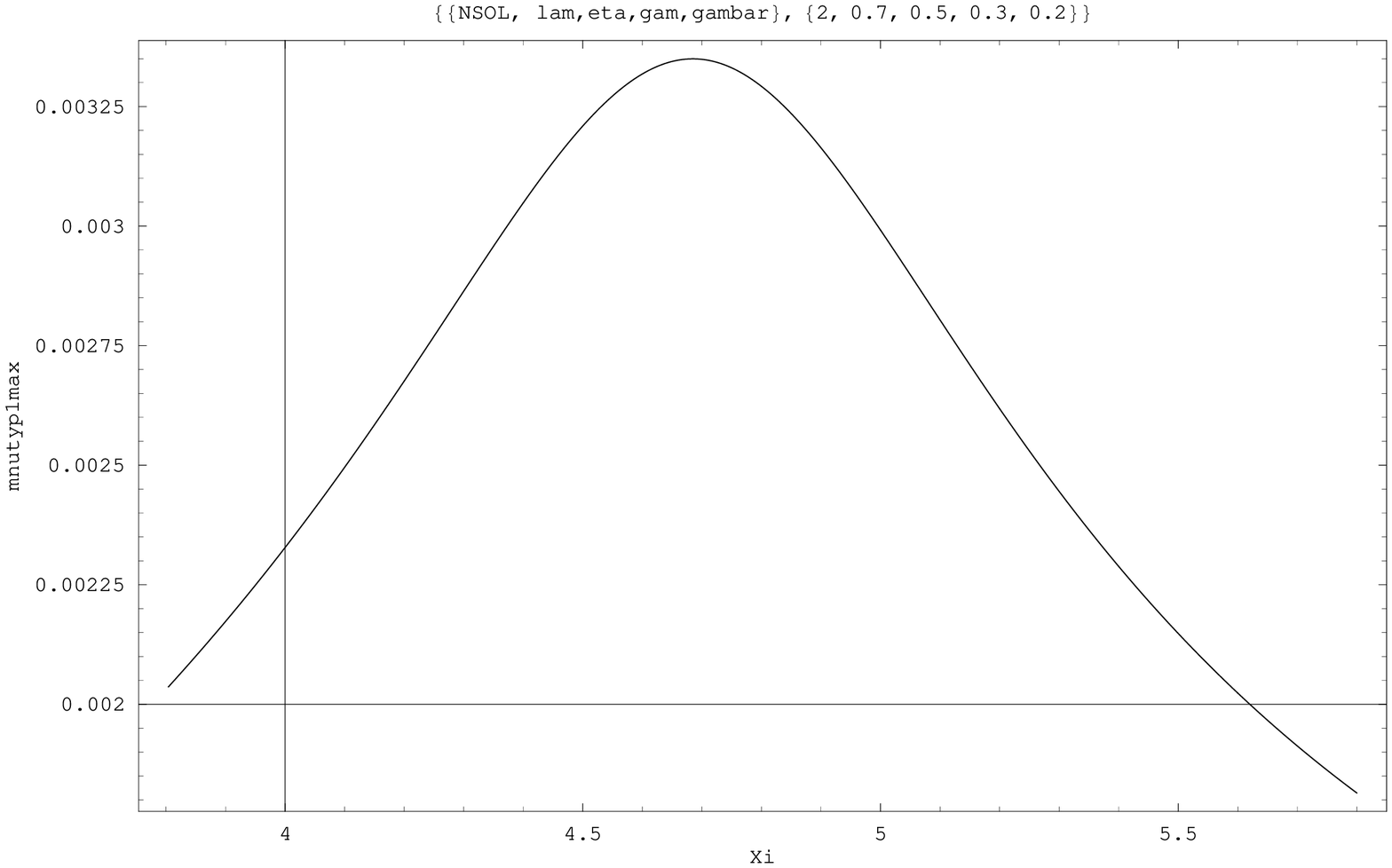}
\epsfxsize15cm\epsffile{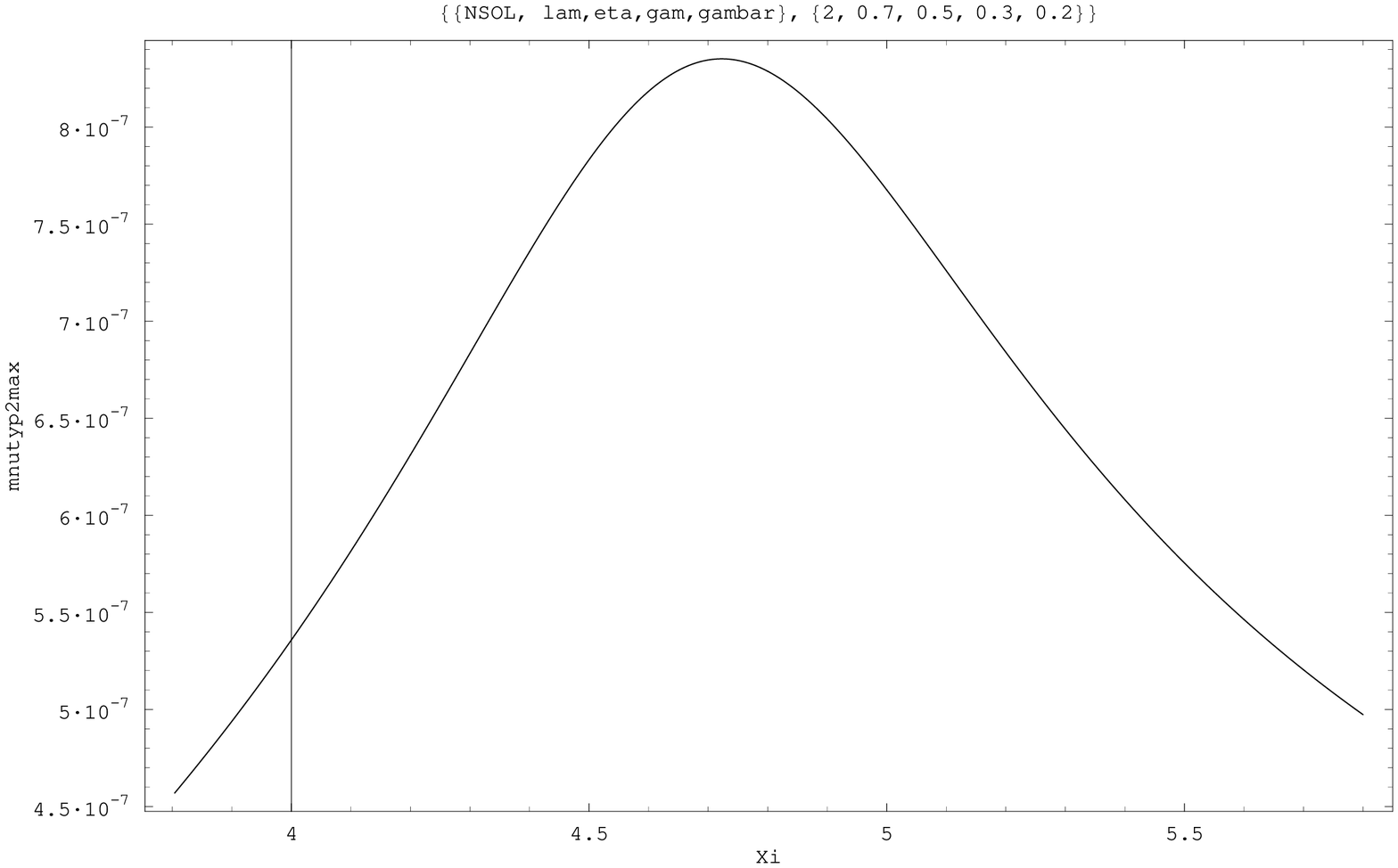}
 \caption{ Magnified plots of  the maximum neutrino masses in the MSGUT
(obtained using the sample solution of Bertolini and Malinsky)   vs $\xi$ for real $\xi$ :
  complex solution   for x,  $\lambda =.7$.  Type I  (upper) and Type II (lower).
Type I solutions  achieve about $10^{-1}$ of the required values at their maximum. }
\end{center}
\end{figure}

\end{document}